%% file: main.tex
\documentclass[fleqn,usenatbib]{mnras}

\usepackage[T1]{fontenc}

\DeclareRobustCommand{\VAN}[3]{#2}
\let\VANthebibliography\thebibliography
\def\thebibliography{\DeclareRobustCommand{\VAN}[3]{##3}\VANthebibliography}

\usepackage{graphicx}	
\usepackage{grffile}    
\usepackage{amsmath}	
\usepackage{amssymb}	
\usepackage{subfiles}
\usepackage{bm}
\usepackage{booktabs}
\usepackage{multirow}
\usepackage{multicol}
\usepackage{subcaption}
\usepackage{xcolor}

\usepackage{txfonts}

\newcommand{\density}{\rho}

\newcommand{\autocovprime}[1]{k_{#1}}
\newcommand{\psdcorrprime}[1]{P_{#1}}

\newcommand{\pdfcovprimenorm}{R}
\newcommand{\densityfluctuations}{\Delta{}\rho}
\newcommand{\gaia}{\emph{Gaia}}

\makeatletter
\providecommand*{\diff}{\@ifnextchar^{\DIfF}{\DIfF^{}}}
\def\DIfF^#1{%
    \mathop{\mathrm{\mathstrut d}}\nolimits^{#1}\gobblespace
}
\def\gobblespace{%
    \futurelet\diffarg\opspace
}
\def\opspace{%
    \let\DiffSpace\!%
    \ifx\diffarg(%
        \let\DiffSpace\relax
    \else
        \ifx\diffarg[%
            \let\DiffSpace\relax
        \else
            \ifx\diffarg\{%
                \let\DiffSpace\relax
            \fi\fi\fi\DiffSpace
}

\newcommand{\E}[1]{\left\langle#1\right\rangle}
\DeclareMathOperator{\var}{var}
\DeclareMathOperator{\cov}{cov}

\DeclareMathOperator{\mse}{MSE}
\DeclareMathOperator{\bias}{bias}
\newcommand{\trans}{\mathrm{t}}
\newcommand{\score}{R_{\mathrm{LOO}}}

{\newif\ifnotend
\notendtrue
\def\veclist{ABCDEFGHIJKLMNOPQRSTUVWXYZabcdfghijklmnopqrstuvwxyz.}
\def\top#1#2.{#1}
\def\tail#1#2.{#2.}
\loop\expandafter\xdef\csname v\expandafter\top\veclist\endcsname%
{{\noexpand\bf\expandafter\top\veclist}}
\edef\veclist{\expandafter\tail\veclist}
\if\veclist.\notendfalse\fi\ifnotend\repeat}

\usepackage{lipsum}

\title[Mapping dust in the giant molecular cloud Orion A]{Mapping dust in the giant molecular cloud Orion A}

\author[A. Gration and J. Magorrian]{
Amery Gration$^{1, 2}$\thanks{E-mail: a.gration@surrey.ac.uk}
and John Magorrian$^{1}$
\\
$^{1}$Rudolf Peierls Centre for Theoretical Physics, Beecroft Building, Parks Road, Oxford, OX1 3PU, UK
\\
$^{2}$Department of Physics, University of Surrey, Guildford, GU2 7XH, UK
}

\date{Accepted XXX. Received YYY; in original form ZZZ}
\pubyear{2023}

\begin{document}
\label{firstpage}
\pagerange{\pageref{firstpage}--\pageref{lastpage}}
\maketitle

\begin{abstract}
  \subfile{abstract.tex}
\end{abstract}

\begin{keywords}
ISM: clouds, ISM: dust, extinction, ISM: structure, Galaxy: local interstellar matter, Galaxy: solar neighbourhood
\end{keywords}


\section{Introduction}

\subfile{introduction.tex}

\section{Statistics of the interstellar medium}

\subfile{statistics_of_the_ism.tex}

\section{Linear prediction and map making}

\subfile{linear_prediction_and_map_making.tex}

\section{Methodological tests using synthetic data}

\subfile{methodological_tests_using_synthetic_data.tex}

\section{Mapping dust in Orion A}

\subfile{mapping_the_ism.tex}

\section{Discussion}

\subfile{discussion.tex}

\section{Conclusion}

\subfile{conclusion.tex}

\section*{Acknowledgements}

\subfile{acknowledgements.tex}

\section*{Data Availability}

\subfile{data_availability.tex}


\bibliographystyle{mnras}
\bibliography{bibliography} 


\appendix

\section{Computing the predicted density}

\subfile{appendix_1.tex}

\section{Prediction intervals and the distribution of the BLUP}

\subfile{appendix_2.tex}

\section{Validation of the maps}

\subfile{appendix_3.tex}

\section{The autocovariance functions}

\subfile{appendix_4.tex}


\bsp	
\label{lastpage}
\end{document}

%% file: abstract.tex
The Sun is located close to the Galactic mid-plane, meaning that we observe the Galaxy through significant quantities of dust.
Moreover, the vast majority of the Galaxy's stars also lie in the disc, meaning that dust has an enormous impact on the massive astrometric, photometric and spectroscopic surveys of the Galaxy that are currently underway.
To exploit the data from these surveys we require good three-dimensional maps of the Galaxy's dust.
We present a new method for making such maps in which we form the best linear unbiased predictor of the extinction at an arbitrary point based on the extinctions for a set of observed stars.
This method allows us to avoid the artificial inhomogeneities (so-called `fingers of God') and resolution limits that are characteristic of many published dust maps.
Moreover, it requires minimal assumptions about the statistical properties of the interstellar medium.
In fact, we require only a model of the first and second moments of the dust density field.
The method is suitable for use with directly measured extinctions, such as those provided by the Rayleigh--Jeans colour excess method, and inferred extinctions, such as those provided by hierarchical Bayesian models like StarHorse.
We test our method by mapping dust in the region of the giant molecular cloud Orion A.
Our results indicate a foreground dust cloud at a distance of 350~pc, which has been identified in work by another author.

%% file: introduction.tex
\label{sec:introduction}

The giant molecular clouds Orion A and B are sites of continuing star formation.
Indeed, they are the regions closest to Earth in which high-mass ($M > 8$~M$_{\sun}$) star formation is occurring.
Both Orion A and B are filamentary structures.
Orion A, which is comet-like, consists of a dense head (containing the Orion Nebula, at galactic coordinates $l = 206$~deg, $b = -16.4$~deg, and its largest cluster of young stars, the Orion Nebula Cluster) and a diffuse tail extending south-west on the sky for several degrees.
Orion B is more uniform.
It contains the Flame Nebula at its eastern extremity (at galactic coordinates $l = 209$~deg, $b = -19.4$~deg) and extends north-west on the sky, again for several degrees.

Photometric surveys of young stars and young stellar objects, which trace the gas of the molecular cloud in which they have been concieved and born, have charted the gross on-sky distribution of Orion A.
These surveys include those by:
\cite{meingast_vision_2016}, \cite{meingast_vision_2018},
and \cite{grosschedl_vision_2019}, as part of the VISION surveys, which used near-infrared observations made by the VISTA telescope augmented by visible and near-infrared observations by Pan-STARRS and mid-infrared observations by Spitzer;
\cite{megeath_spitzer_2012, megeath_spitzer_2015}, which used mid-infrared observations made by Spitzer and near-infrared observations made by 2MASS;
\cite{carpenter_2mass_2000}, which used near-infrared observations made by 2MASS; and
\cite{wright_wide-field_2010}, which used mid-infrared observations by WISE.
Astrometric surveys of the same young stars and young stellar objects have allowed us to chart the gross three-dimensional distribution of Orion A.
These surveys include those by
\cite{kounkel_apogee-2_2018} and \cite{grosschedl_3d_2018}, using \emph{Gaia} Data Release 2 (DR2).
They have established that the head of Orion A is a roughly spherical object, with a diameter of $15$ to $20$~pc, lying at a distance of $s = 400$~pc at $l = 209~\mathrm{deg}$, $b = -19.5~\mathrm{deg}$, and that its tail extends away from the Sun for a distance of $75$~pc at an angle of $70$~deg to the plane of the sky.

To chart the fine three-dimensional distribution of Orion A we can use dust, which is coupled to the gas component of the interstellar medium (ISM) and hence traces the densest gas of the molecular cloud in which it sits.
The dust is accessible to us by its extincting effect on light passing through it.
But the task of charting the dust of Orion A is very different from the task of charting its stellar population.
It is a matter of inference rather than direct observation.
Given extinctions for stars in the region of Orion A we can \emph{infer} the dust density at an arbitrary point in the cloud, a process known as `three-dimensional dust mapping'.
A number of teams have already made three-dimensional dust maps of Orion A.
These include:
\cite{schlafly_three-dimensional_2015}, using only photometry from Pan-STARRS1;
\cite{rezaei_three-dimensional_2018}, using astrometry from \emph{Gaia} Data Release 1 together with photometry from 2MASS and WISE,
\cite{rezaei_detailed_2020}, using astrometry from \emph{Gaia} DR2 together with photometry from 2MASS and WISE; and \cite{dharmawardena_three-dimensional_2022}, using astrometry from \emph{Gaia} DR2 together with photometry from 2MASS, WISE, and \emph{Gaia} DR2.
All three identify a filament of gas consistent with the tail of young stellar objects and young stars.
\cite{rezaei_detailed_2020} report a cloud lying in the foreground of Orion A, at a distance of $s = 350$~pc, which they tentatively associate with a stellar cluster identified by \citet{bouy_orion_2014}.
However, this cloud is not reported by either of the other two groups.

Although the distributions of the gas and dust are of interest in their own right, the dust is also something of a nuisance since it obscures and reddens observations of stars within and beyond it.
Because the Sun is located close to the Galactic mid-plane we observe our galaxy through significant quantities of dust.
Moreover, the vast majority of the Galaxy's stars also lie in the disc, so extinction has an enormous impact on stellar surveys and, in turn, on attempts to fit chemical and dynamical models of the Galaxy to stellar observations.
To fully exploit the data from stellar surveys we require good three-dimensional dust maps of the whole Galaxy.
A number of teams have made such three-dimensional dust maps.
These include:
\cite{arenou_tridimensional_1992};
\cite{drimmel_three-dimensional_2003};
\cite{marshall_modelling_2006};
\cite{sale_3d_2014} (based on the method of \citealp{sale_3d_2012});
\cite{chen_three-dimensional_2014} (based on the method of \citealt{berry_milky_2012});
\cite{lallement_3d_2015}, \cite{lallement_3d_2014}, \cite{lallement_three-dimensional_2018}, \cite{lallement_gaia-2mass_2019}, and \cite{vergely_spatial_2010} (based on the method of \citealp{vergely_nai_2001});
\cite{green_three-dimensional_2015}, \cite{green_galactic_2018}, and \cite{green_3d_2019} (based on the method of \citealp{green_measuring_2014}); and
\cite{leike_charting_2019} and \cite{leike_resolving_2020} (based on the method of \cite{enslin_inference_2010} and \citealp{enslin_reconstruction_2011}).

To date, however, dust maps have suffered from either artificial inhomogenities known as `fingers of God'
\citep{marshall_modelling_2006, green_galactic_2018, lallement_gaia-2mass_2019}, or
have been resolution-limited
\citep{sale_3d_2014, rezaei_detailed_2020, leike_resolving_2020}.
We propose a new method for creating three-dimensional dust maps and use it to map Orion A.
Orion A is a good testing ground for dust-mapping methods that might be used to map the whole Galaxy.

When making a three-dimensional dust map we wish to infer the density, $\rho(\bm{r})$ or, equivalently, the extinction, $A(\bm{r})$, at an arbitrary point, $\bm{r}$, given knowledge of the positions, $\bm{r}_{1}, \dots , \bm{r}_{n}$, and extinctions, $A_{1}, \dots , A_{n}$ for some set of stars.
In this paper we treat the ISM as the realization of a random field, and compute the best linear unbiased predictor of the extinction at a point given extinctions for a set of stars in the region of that point.
This is equivalent to forming the generalized least squares estimator of the mean of the extinction and then performing a nonparameteric fit to the residuals.
We then differentiate this predictor to give a predictor of the density at the same point. 
Our method requires minimal assumptions about the statistical properties of the ISM.
Indeed it requires us to know only the covariance of the densities of the ISM and to have a model of its mean.
It does not require us to assume a distribution for the dust density, and in particular it does not require us to assume a log-normal distribution, as is common in three-dimensional dust mapping.
We follow \cite{sale_three-dimensional_2014} in using a physically motivated model of the covariance function that captures the turbulent structure of the ISM.
The parameters of this covariance model form the hyperparameters of our predictor of extinction, which we optimize using the method of leave-one out cross validation.
We make two maps of dust in the region of Orion A out to $500$~pc: one using extinctions computed using the Rayleigh--Jeans colour excess method, and one using extinctions from the StarHorse catalague.
The resulting maps display no fingers of God, and may be constructed at arbitary resolution although the effective resolution is always limited by the density of observed stars.

Our method relies on knowledge of only the broadest statistical properties of the density field (i.e. its first and second moments) and we begin, in Section~\ref{sec:statistics_of_the_ism}, by summarizing these.
We then introduce the best linear unbiased predictor, in Section~\ref{sec:linear_prediction_and_map_making}, testing it against simulated data in Section~\ref{sec:methodological_tests_using_synthetic_data} before using it to map Orion A, in Section~\ref{sec:mapping_the_ism}.
Our maps broadly agree with those made by \cite{rezaei_detailed_2020} and, like them, show a foreground dust cloud at a distance of 350 pc.
In Section~\ref{sec:discussion} we discuss the consequences of our assumptions about the statistical properties of the density field and show that our method is insensitive to them so long as they correctly account for the fractal nature of the ISM.

%% file: statistics_of_the_ism.tex
\label{sec:statistics_of_the_ism}

The density of the ISM at a point, $\bm{r}$, may be represented by the random variable, $\density(\bm{r})$.
The set of all such random variables forms a random field, $\density := (\density(\bm{r}))_{\bm{r} \in \mathbf{\bf{R}}^{3}}$, which represents the density of the ISM as a whole.
Similarly, the extinction of light emitted by a source at a point, $\bm{r}$, may be represented by the random variable
\begin{align}
    A(\bm{r}) = \int_{0}^{s}\density(s'\hat{\bm{r}})\diff{}s'
\end{align}
where $s := |\bm{r}|$ is the line-of-sight distance of the source and $\hat{\bm{r}}$ is the unit vector parallel to $\bm{r}$.
This gives another random field, $A := (A(\bm{r}))_{\bm{r} \in {\bf{}R}^{3}}$, which represents the extinction that would be undergone by light emitted at any point in space.
The properties of these fields are usefully summarized by their means and variances. 
Since the expection and covariance functions are both linear we immediately find that the expectation of $A(\mathbf{r})$
is
\begin{align}
  \E{A(\bm{r})}
  &= \E{\int_{0}^{s}\density(s'\hat{\bm{r}})\diff{}s'}\\
  &= \int_{0}^{s}\E{\density(s'\hat{\bm{r}})}\diff{}s'
  \label{eq:ext_expectation}
\end{align}
and that the covariance of $A(\mathbf{r})$ is%
\footnote{What \citet[][eq.~A2]{sale_three-dimensional_2014} call the correlation, $\E{A(\bm{s}_{1})A(\bm{s}_{2})}$, should be the covariance, $\cov(A(\bm{s}_{1}), A(\bm{s}_{2}))$.}
\begin{align}
  \cov(A(\bm{r}_{i}), A(\bm{r}_{j}))
  &= \cov\left(
    \int_{0}^{s_{i}}\density(s'_{i}\hat{\bm{r}}_{i})\diff{}s'_{i},
    \int_{0}^{s_{j}}\density(s'_{j}\hat{\bm{r}}_{j})\diff{}s'_{j},
    \right)\\
  &= \int_{0}^{s_{j}}\int_{0}^{s_{i}}\cov(\density(s'_{i}\hat{\bm{r}}_{i}), \density(s'_{j}\hat{\bm{r}}_{j}))\diff{}s'_{i}\diff{}s'_{j},
\label{eq:ext_covariance}
\end{align}
where $s_{i} := |\bm{r}_{i}|$ and $s_{j} := |\bm{r}_{j}|$ are the line-of-sight distances of the two sources and $\hat{\bm{r}}_{i}$ and $\hat{\bm{r}}_{j}$ are the unit vectors parallel to $\bm{r}_{i}$ and $\bm{r}_{j}$. Although the ISM is a complicated multiphase medium, the fluctuation in its density,
\begin{align}
    \label{eq:density_fluctuations}
    \densityfluctuations := \density - \E{\density},
\end{align}
can be approximated as being stationary and isotropic \citep{draine_physics_2011}.
In these circumstances, its first and second moments are invariant under translations and rotations.
Consequently, the covariance of the density fluctuations at two points, $\Delta\rho(\bm{r}_{i})$ and $\Delta\rho(\bm{r}_{j})$, is a function of their separation, $|\bm{r}_{j} - \bm{r}_{i}|$, and is given by the \emph{autocovariance function} of the density fluctuations, $\autocovprime{\densityfluctuations}$, such that
\begin{align}
  \autocovprime{\densityfluctuations}(|\bm{r}_{j} - \bm{r}_{i}|)
  &= \cov(\densityfluctuations(\bm{r}_{i}), \densityfluctuations(\bm{r}_{j})).
\end{align}
By Bochner's Theorem \citep{adler_geometry_1981} the Fourier transform of $\autocovprime{\densityfluctuations}$ exists, and is related to the \emph{power spectrum} of $\Delta{}\rho$, denoted $\psdcorrprime{\densityfluctuations}$, via the equation
\begin{align}
    \label{eq:autocorrelation_model}
    \autocovprime{\densityfluctuations}(|\bm{r}_{j} - \bm{r}_{i}|) &= \int_{{\bf{}R}^{3}}\mathrm{e}^{\mathrm{i}{}\bm{k}\cdot{}(\bm{r}_{j} - \bm{r}_{i})}\psdcorrprime{\densityfluctuations}(|\bm{k}|)\diff{}|\bm{k}|,
\end{align}
where $\bm{k}$ is the three-dimensional wavenumber.

It is common to assume that fluctuations in the density are described by Kolmogorov's theory of turbulence \citep{draine_physics_2011}, i.e.\ that its power spectrum is described by a power law, with index $-11/3$, over a wide range of wavenumbers.%
\footnote{%
This is a remarkable fact since Kolmogorov's theory of tubulence describes the behaviour of small-amplitude incompressible hydrodynamic systems, and the ISM is a large-amplitude compressible magnetohydrodynamic system \citep{draine_physics_2011}.
}
This power law must fail at very small and very large wavenumbers.
In the Kolmogorov theory of turbulence this happens at the boundaries of the inertial regime, i.e.\ at scales larger than the energy-injection scale, $L_{0}$, or smaller than the dissipation scale.
The energy-injection scale is that of stellar feedback, while the energy-dissipation scale is that of viscous forces.
Although astronomical observations clearly probe the energy-injection scale, they do not probe the energy-dissipation scale, meaning that we need only worry about the failure of the power-law power spectrum for small wavenumbers.
Motivated by this fact, \cite{sale_three-dimensional_2014} proposed a split-power law for the power spectrum of $\densityfluctuations$, with the location of the break being determined by the energy-injection scale, $L_{0}$.
It is given by
\begin{align}
  \label{eq:psd_model}
  \psdcorrprime{\densityfluctuations}(|\bm{k}|) = \sigma_{\densityfluctuations}^{2}\pdfcovprimenorm\dfrac{(|\bm{k}|L_{0})^{2\Omega}}{(1 + (|\bm{k}|L_{0})^2)^{\gamma/2 + \Omega}}
\end{align}
for positive numbers $\Omega$, $\gamma$, and $\pdfcovprimenorm$, a normalizing constant defined implicitly by the equation
\begin{align}
    \dfrac{1}{\pdfcovprimenorm} = 4\pi\int_{0}^{\infty}\dfrac{|\bm{k}|^{2}(|\bm{k}|L_{0})^{2\Omega}}{(1 + (|\bm{k}|L_{0})^2)^{\gamma/2 + \Omega}}\diff{}|\bm{k}|.
\end{align}
Above the wavelength associated with the energy-injection scale, $L_{0}$, the spectrum is a power-law with index $-\gamma$, and below that wavelength it is a power law with index $\Omega$.
\citeauthor{sale_three-dimensional_2014} call this power spectrum `Kolmogorov-like', regardless of the value of $\gamma$.

The mean value of the density is in general unknown, but its covariance is equal to the covariance of the density fluctuations, since
\begin{align}
  \cov(\density(\bm{r}_{i}), \density(\bm{r}_{j}))
  &= \E{(\density(\bm{r}_{i}) - \E{\density(\bm{r}_{i})})(\density(\bm{r}_{j}) - \E{\density(\bm{r}_{j})})}\\
  &= \cov(\densityfluctuations(\bm{r}_{i}), \densityfluctuations(\bm{r}_{j})).
\end{align}
The density, $\density$, is often taken to be lognormal, meaning that the extinction at a point, $A(\bm{r})$, is the integral of a lognormal field.
This has a probability density function (PDF) that cannot be expressed in closed form, but which is itself approximately lognormal \citep{ostriker_density_2001}.

If we assume that the density (as well as its fluctuations)%
\footnote{
  Of course the density of the ISM decreases exponentially as Galactic radius increases. Morever, at large scales, the ISM forms complex sheet-like and filamentary structures. At these large scales the assumptions of stationary density and stationary density fluctuations are poor. However, at the small scales we consider in this paper both assumptions are reasonable.
}
is stationary and isotropic, with mean $\mu_{\density}$ and constant variance $\sigma^{2}_{\density}$ then the covariance of the densities at two points, $\rho(\bm{r}_{i})$ and $\rho(\bm{r}_{j})$, is also a function of their separation, $|\bm{r}_{j} - \bm{r}_{i}|$, and is given by the autocovariance function of the density, $\autocovprime{\density}$, such that 
\begin{align}
  \autocovprime{\density}(|\bm{r}_{j} - \bm{r}_{i}|) = \cov(\rho(\bm{r}_{i}), \rho(\bm{r}_{j})).
\end{align}
The autocovariance of the density is then equal to the autocovariance of the density fluctuations, i.e.\ $\autocovprime{\density} = \autocovprime{\densityfluctuations}$, and we have that
\begin{gather}
  \label{eq:mean_density}
  \E{A(\bm{r})} = \mu_{\density}s,\\
  \label{eq:cov_ext}
  \cov(A(\bm{r}_{i}), A(\bm{r}_{j}))
  = \int_{0}^{s_{i}}\int_{0}^{s_{j}}\autocovprime{\densityfluctuations}(|s'_{i}\hat{\bm{r}}_{i} - s'_{j}\hat{\bm{r}}_{j}|)\diff{}s'_{j}\diff{}s'_{i}
\end{gather}
where $\autocovprime{\densityfluctuations}$ is given by Equations~\ref{eq:autocorrelation_model} and~\ref{eq:psd_model}.
For each choice of autocovariance function, we need only a model of the mean of the extinction, which has a single free parameter, $\mu_{\density}$.
Despite this simplifying assumption, neither $\cov(A(\bm{r}_{i}), A(\bm{r}_{j}))$ nor $\cov(\density(\bm{r}_{i}), A(\bm{r}_{j}))$ has closed form, and both must be computed numerically at some expense.
For this reason we define the functions $f$ and $g$ such that%
\begin{align}
  \label{eq:function_f}
  f(s_{1}, s_{2}, \theta_{12}) &= \cov(\density(\bm{r}_{1}), \density(\bm{r}_{2}))\\
  \label{eq:function_g}
  g(s_{1}, s_{2}, \theta_{12}) &= \cov(A(\bm{r}_{1}), A(\bm{r}_{2}))
\end{align}
where $s_{1}$ and $s_{2}$ are the line-of-sight distances to points $\bm{r}_{1}$ and $\bm{r}_{2}$, and $\theta_{12}$ is the angle subtended by them at the origin, which we precompute on a regular lattice of arguments and evaluate using linear interpolation. (We discuss these functions further in App.~\ref{sec:appendix_4}).

%% file: linear_prediction_and_map_making.tex
\label{sec:linear_prediction_and_map_making}

When making a three-dimensional dust map we wish to infer the density or, equivalently, the extinction at an arbitrary point given knowledge of the extinctions and positions for some set of stars.
Of course, neither the distance nor the extinction are observed directly.
Instead, they must themselves be inferred from other stellar properties.
For a limited number of stars (those for which we have observations in the infrared) we may estimate the extinction using the near-infrared colour excess (NICE) method of \cite{lada_dust_1994} or its successors
(the NICER method due to \citealt{lombardi_mapping_2001},
the NICEST method due to \citealt{lombardi_nicest_2009}, and
the Rayleigh--Jeans colour excess method due to \citealt{majewski_lifting_2011}).
The Rayleigh--Jeans colour excess (RJCE) method, for example, exploits the fact that all stars are well approximated as black bodies in the low-frequency limit (where their spectral radiance is described by the Rayleigh--Jeans law), meaning that their low-frequency colours are approximately independent of spectral class. \cite{majewski_lifting_2011} showed that colours spanning the NIR and MIR are approximately constant for a range of spectral types, and that the colour $H - [4.5~\mu]$ exhibits especially little variation.
They found, using the extinction law of \cite{cardelli_relationship_1989}, that the $K_{\mathrm{s}}$-band extinction in magnitudes is
\begin{align}
  A_{K_{\mathrm{s}}} = 0.918(H - [4.5~\mu] - 0.08)
\end{align}
with a scatter of 0.1 mag for F, G, and K stars, and 0.4 mag for B to M excluding late type dwarfs.
However, due to observational error, both distances and extinctions estimated in this way may be unphysically negative.

However, it is often necessary to infer the extinction and distance of a star using Bayesian methods of the kind pioneered by \cite{burnett_stellar_2010}, \cite{bailer-jones_bayesian_2011}, and \cite{binney_galactic_2014}.
These methods assume that a star's observed properties are parameterized by its instrinsic properties, and use Bayes's theorem to compute the posterior PDF of the intrinsic properties given the observed properties.
We might wish to infer a single property or a tuple of properties, in which case we can find the marginal PDF of any element of that tuple.
Several catalogues of stellar properties have been compiled in this way.
They include %
the StarHorse catalogue (\citealt{santiago_spectro-photometric_2016}, \citealt{queiroz_starhorse_2018}, \citealt{anders_photo-astrometric_2019}, \citealt{anders_photo-astrometric_2021}), and %
the catalogues of Bailer-Jones and his collaborators
(\citealt{bailer-jones_estimating_2015},
\citealt{astraatmadja_estimating_2016a, astraatmadja_estimating_2016b}, and
\citealt{bailer-jones_estimating_2018, bailer-jones_estimating_2021}).

The StarHorse catalogue takes the intrinsic properties of a star to be mass, age, distance, and $V$-band extinction.
It takes the observable properties of a star to be effective temperature, surface gravity, overall metallicity, magnitude, and parallax, as catalogued by \emph{Gaia}, Pan-STARRS1, 2MASS, and AllWISE.
The likelihood of these observable properties is assumed to be Gaussian with mean given by a stellar model, and variance given by the observational errors.
It assumes as a prior PDF for distance a physically motivated four-component model of the Galaxy (comprising the thin disc, thick disc, bulge/bar and halo). As a result, inferred distances are always positive.
In the first release of the StarHorse catalogue, using \emph{Gaia} DR2 \citep{anders_photo-astrometric_2019}, the prior PDF for extinctions is assumed to be uniform in the case of stars with low-noise parallax observations (signal-to-noise of five or greater) and a top hat, such that $A_{V}/\text{mag} \in [-0.3, 4]$, in the case of stars with high-noise parallax observations (signal-to-noise of less than five).
In the second release, using \emph{Gaia} Early Data Release 3 (EDR3) \citep{anders_photo-astrometric_2021}, the prior PDF for extinctions is assumed to be given by the maps of \cite{green_3d_2019} or \cite{drimmel_three-dimensional_2003} according to coverage.
The full posterior PDF is not publicly available, but rather the $0.05$, $0.16$, $0.6$, $0.84$, and $0.95$ quantiles.

The catalogues of Bailer-Jones provide only distances.
They take the instrinsic property of a star to be distance, and its observable property to be parallax.
The most recent of these catalogues \citep{bailer-jones_estimating_2018, bailer-jones_estimating_2021} assumes that the likelihood of the parallax is Gaussian with mean given by the reciprocal distance and variance given by the observational errors, and that the prior PDF for distance is that for a generalized gamma distribution.
As a result, inferred distances are again always positive.
Again, the full posterior PDF is not publicly available, but rather the $0.05$, $0.16$, $0.6$, $0.84$, and $0.95$ quantiles.

\subsection{The best linear unbiased predictor}

For a list of directly computed or inferred extinctions
$\bm{A} = (A_{1}, \dots , A_{n})$ at locations
$\bm{r}_{1}, \dots , \bm{r}_{n}$
we wish to predict the extinction, $A(\bm{r})$, or density, $\rho(\bm{r})$, at some arbitrary position, $\bm{r}$, given the assumptions represented by Eqs~\ref{eq:mean_density} and~\ref{eq:cov_ext}.
To predict the extinction we will use the \emph{best linear unbiased predictor} (BLUP) of $A(\bm{r})$, which we will denote $\hat{A}(\bm{r})$.
This is a linear combination of extinctions $A_{1}, \dots , A_{n}$ with coefficients chosen so as to minimize the mean-square error,
\begin{align}
  \operatorname{MSE}(\hat{A}(\bm{r})) = \E{(A(\bm{r}) - \hat{A}(\bm{r}))^{2}}
\end{align}
(where angle brackets again indicate the expectation) subject to the constraint that $\hat{A}(\bm{r})$ be unbiased, i.e.\ subject to the constraint that
$\E{\smash{\hat{A}(\bm{r})}} = \E{A(\bm{r})}$.
It is given by \cite{goldberger_best_1962} as
\begin{align}
  \label{eq:blup}
  \hat{A}(\bm{r})
  = \bm{\gamma}(\bm{r})^{\trans}\bm{A}
  + \bm{\sigma}(\bm{r})^{\trans}\bm{\Sigma}^{-1}(\bm{A} - \bm{\Gamma}^{\trans}\bm{A})
\end{align}
where the $n \times 1$ vector-valued function $\bm{\gamma}$ is given by
\begin{align}
  \bm{\gamma}(\bm{r}) = \bm{\Sigma}^{-1}{\bf{}\Phi}^{\trans}({\bf{}\Phi}\bm{\Sigma}^{-1}{\bf{}\Phi}^{\trans})^{-1}s,
\end{align}%
the $n \times 1$ vector-valued function $\bm{\sigma}$ is given element-wise by
$[\bm{\sigma}(\bm{r})]_{i} = \cov(A(\bm{r}), A_{i})$,
the $n \times n$ matrix $\bm{\Sigma}$ has elements
$[\bm{\Sigma}]_{ij} = \cov(A_{j}, A_{i})$,
the $n \times n$ matrix $\bm{\Gamma}$ has columns $[\bm{\Gamma}]_{j} = \bm{\gamma}(\bm{r}_{j})$, meaning that it has elements
$[\bm{\Gamma}]_{ij} = [\bm{\gamma}(\bm{r}_{j})]_{i}$, and
the $n \times 1$ vector $\bm{\Phi}$ has elements
$[\bm{\Phi}]_{i} = s_{i}$.

The expression for $\hat{A}(\bm{r})$ is the sum of two terms.
The first of these, $\bm{\gamma}(\bm{r})^{\trans}\bm{A}$, is the generalized least squares (GLS) estimator of the mean of $A(\bm{r})$, which we may rewrite as $\hat{\mu}_{\density}s$, where
\begin{align}
  \hat{\mu}_{\density} := (\bm{\Phi}^{\trans}\bm{\Sigma}^{-1}\bm{\Phi})^{-1}\bm{\Phi}\bm{\Sigma}^{-1}\bm{A}
\end{align}
is itself the GLS estimator of the mean of the density, $\mu_{\density}$, and is unbiased.
Similarly, the expression $\bm{\Gamma}^{\trans}\bm{A}$ gives the GLS estimators of the means of $A_{1}, \dots , A_{n}$. The second term in the expression for $\hat{A}(\bm{r})$ is then a weighted sum of the residuals of the observed values $\bm{A}$ and the GLS of their means.%
\footnote{A similar method of prediction, known as `kriging' is much used in the mining industry for mapping geological features \citep[see, for example,][]{cressie_statistics_1993}.}

To predict the density we will use the derivative of the BLUP of $A(\bm{r})$,
\begin{align}
  \label{eq:predictor_density}
  \hat{\density}(\bm{r})
  := \dfrac{\partial{}\hat{A}(\bm{r})}{\partial{}s},
\end{align}
which is also unbiased, since
$\E{\density(\bm{r}) - \hat{\density}(\bm{r})}
= \E{\smash{{\partial{}}(A(\bm{r}) - \hat{A}(\bm{r}))/{\partial{}s}}}
= {\partial{}}\E{\smash{A(\bm{r}) - \hat{A}(\bm{r})}}/{\partial{}s}
= 0$.
Because only $\bm{\gamma}$ and $\bm{\sigma}$ are functions of $s$ it is easy to compute (see App.~\ref{sec:appendix_1} for details).

Note that to find $\hat{A}(\bm{r})$ and $\hat{\rho}(\bm{r})$ we need only to know the covariance of $(A(\bm{r}), \bm{A})$ and to assume that the expectation of $A(\bm{r})$ is linear in $s$.
We do not need to make any further assumptions about the distribution of the ISM.
In particular, we do not need to assume that its distribution is lognormal.%
\footnote{
Our expression for $\hat{A}(\bm{r})$ is identical to that used in Gaussian-process emulation, where it would be derived under the assumption that $A(\bm{r})$ and $\bm{A}$ are drawn from an underlying Gaussian random field \citep[e.g.][]{rasmussen_gaussian_2006}.
However this assumption of Gaussianity is unnecessary.
The expression holds because the random field $A$ is second order, not because $A$ is Gaussian (which it need not be).
}

\subsubsection{Computing the BLUP in practice}
\label{sec:computing_the_blup}

When using extinctions inferred with the RJCE method we can make the decomposition $A_{i} = A(\bm{r}_{i}) + E_{i}$ where $A(\bm{r}_{i})$ is the intrinsic extinction at $\bm{r}_{i}$ and $E_{i}$ is the uncertainty in its value.
In this case the covariance matrix is
\begin{align}
    [\bm{\Sigma}]_{ij}
    &= \cov(A(\bm{r}_{i}) + E_{i}, A(\bm{r}_{j}) + E_{j})\\
    \label{eq:cov_matrix}
    &= \cov(A(\bm{r}_{i}), A(\bm{r}_{j})) + \var(E_{i})\delta_{ij},
\end{align}
assuming that $A(\bm{r}_{i})$ and $E_{j}$ are independent for $i \neq j$, and
\begin{align}
    [\bm{\sigma}(\bm{r})]_{i}
    &= \cov(A(\bm{r}), A(\bm{r}_{i}) + E_{i})\\
    \label{eq:cov_vector}
    &= \cov(A(\bm{r}), A(\bm{r}_{i})),
\end{align}
assuming that $A(\bm{r})$ and $E_{j}$ are independent.
When using extinctions inferred by Bayesian techniques that make stronger assumptions about he stars' neighbourhoods $A_{i}$ is the posterior prediction of a source's extinction, and both $\cov(A_{i}, A_{j})$ and $\cov(A(\bm{r}), A_{i})$ are properties of the Bayesian model used to construct it.
Since neither is available as part of a catalogue we will approximate them by Equations~\ref{eq:cov_matrix} and~\ref{eq:cov_vector}, on the understanding that $\var(E_{i})\delta_{ij}$ no longer represents an observational error but instead quantifies the discrepancy between our approximation of the covariance and its true value.
We will take $\var(E_{i})$ to be equal to the variance of $\var(A_{i})$.

To compute $\hat{A}(\bm{r})$ we must know the position of each star.
However, as we have already noted, the distance to a star is inherently uncertain.
To circumvent this problem we may assume that the distance to a star is certain but that the uncertainty in the extinction is increased, using the fact that
\begin{align}
    \diff{}A_{i} = \dfrac{\partial{}A_{i}}{\partial{}s_{i}}\Delta{}s_{i}.
\end{align}
In this case, the variance of $A_{i}$ is increased by $\diff{}A_{i}^{2}$.
Under this assumption the $ij$-th element covariance matrix, $[\bm{\Sigma}]_{ij}$, becomes $[\bm{\Sigma}]_{ij} + \diff{}A_{i}^{2}\delta_{ij}$,
i.e.\ we add the square of the uncertainties, $\diff{}A_{1}, \dots , \diff{}A_{n}$, to the diagonal elements of $\bm{\Sigma}$.
If $\diff{}s_{i} \ll L_{0}$ then we may make the linear approximation,
\begin{align}
    \dfrac{\partial{}A_{i}}{\partial{}s_{i}} = \dfrac{A_{i}}{s_{i}}.
\end{align}
Of course, if the condition $\diff{}s_{i} \ll L_{0}$ is not met then this additional term in the variance will be too great.

\subsection{Validation}
\label{sec:validation}

Although our statistical model of the ISM is physically motivated, we would still like reassurance that $\hat{A}(\bm{r})$ and $\hat{\density}(\bm{r})$ are good predictors of $A(\bm{r})$ and $\density(\bm{r})$.
In the case of extinction we may test the performance of the BLUP using leave-one-out cross-validation (LOOCV).
We partition $\bm{A}$ into a set containing a single element, $A_{i}$, and another containing $n - 1$ elements,
$\bm{A}_{-i}
:= (
A_{1},
\dots ,
A_{i - 1},
A_{i + 1},
\dots ,
A_{n}
)$.
We then find $\hat{A}(\bm{r}_{i})$, namely the BLUP of $A(\bm{r}_{i})$ based on $\bm{A}_{-i}$, and compute the residual,
$A_{i} - \hat{A}(\bm{r}_{i})$.
The behaviour of these $n$ residuals should be consistent with the assumptions we have made about their distribution.
Since we have only assumed that the mean is linear and that we know the autocovariance function, we need only check that the residuals obey Chebyshev's inequality,
\begin{align}
\operatorname{Pr}(|A(\bm{r}) - \hat{A}(\bm{r})| \ge \lambda{}\operatorname{std}(A(\bm{r}) - \hat{A}(\bm{r})) \le \dfrac{1}{\lambda^{2}}
\end{align}
where $\operatorname{std}(A(\bm{r}) - \hat{A}(\bm{r}))$ is the standard deviation of the residuals and $\lambda$ is a positive number, although we would hope for them to do considerably better.
They should be uniformly good across the mapped region, and display no trend in distance, latitude or longitude.
However, we always expect our predictors to underperform near the boundaries of the mapped region, where they are constrained by fewer data.
It is also useful to define the LOOCV score,
\begin{align}
    \score := \dfrac{1}{n}\sum_{i = 1}^{n}(A_{i} - \hat{A}(\bm{r}_{i}))^{2},
    \label{eq:loocv_score}
\end{align}
which should be small compared to, say, the mean variance of the elements of $\bm{A}$.
Since we are unable to validate the BLUP of the density of the ISM in this way, we will take validation of $\hat{A}(\bm{r})$ to be validation of $\hat{\density}(\bm{r})$.

\subsection{Choosing the model parameters}
\label{sec:choosing_the_model_parameter}

We have observed that for a given autocovariance function, our model of the mean of the extinction has a single free parameter, $\mu_{\density}$.
However, we do not in practice know the values of the parameter tuple that specifies the power spectrum, and hence the autocovariance function.
We will assume that $\gamma = 11/3$ (to ensure Kolmogorov turbulence in the inertial regime), and that $\Omega = 1$ (arbitrarily).
However the variance of the density, $\sigma_{\densityfluctuations}^{2}$, and the length scale, $L_{0}$, are less certain.
We might wish to find their maximum-likelihood estimate (MLE).
But this would require us to assume a model for the distribution of the extinction, contrary to the assumptions of the BLUP, so that by doing this we would lose one of the principal benefits of the method.
Moreover, in computing the BLUP of the extinction we are necessarily computing the GLS estimator of the mean, and we do not want a second estimate of it.
Instead, we may follow a procedure much used in machine learning, and choose the pair $\sigma_{\densityfluctuations}^{2}$ and $L_{0}$ so as to minimize the LOOCV score, $\score$ (eq.~\ref{eq:loocv_score}).
This does not allow us to formally estimate the parameter tuple in the way the method of maximum likelihood does.
We should, instead, see it as a way of tuning the predictor so that it passes validation.
Ultimately, our trust in the BLUP is justified only by its performance in validation, and we may choose the power spectrum's parameter tuple arbitrarily, so long as the resulting predictors pass this test.

\subsection{Computational expense}

The memory requirements for computing the BLUP are dominated by the storage of square matrices of size $n \times n$ and are therefore of order $O(n^{2})$.
The computational complexity is dominated by the inversion of the covariance matrix, $\bm{\Sigma}$.
For small data sets, like those needed to map the dust in Orion A, we may do this using Cholesky decomposition, which has computational complexity of order $O(n^{3})$.
However, for significantly larger data sets, like those needed to create global dust maps, this complexity is prohibitive.
Such scaling problems are common to all three-dimensional dust mapping methods and a number of solutions are available.
In our case it is best to use a low-rank or sparse approximation to the covariance matrix or to use the iterative inversion method described by \cite{wang2019}, which has complexity of order $O(n^{2})$.
We will pursue this in future work.

%% file: methodological_tests_using_synthetic_data.tex
\label{sec:methodological_tests_using_synthetic_data}

\newcommand{\varsynthetic}{\ensuremath{3.0 \times 10^{-6}~\mathrm{mag}^{2}\mathrm{pc}^{-2}}}
\newcommand{\scalesynthetic}{\ensuremath{98~\mathrm{pc}}}
\newcommand{\rsssynthetic}{\ensuremath{0.003~\mathrm{mag}^{2}}}
\newcommand{\bluesynthetic}{\ensuremath{1.2 \times 10^{-3}~\mathrm{mag}\mathrm{pc}^{-1}}}
\newcommand{\bluecisynthetic}{\ensuremath{[8.94 \times 10^{-4}, 1.547 \times 10^{-3}]}}
\newcommand{\rmsesynthetic}{\ensuremath{3.36 \times 10^{-4}}}

We will shortly use our method to map dust within Orion A and its environment.
In fact, we will follow \cite{rezaei_detailed_2020} and \cite{rezaei_three-dimensional_2022} by mapping the region $s/\text{pc} \in [0, 500]$, $l/\text{deg} \in [205, 216]$, $b/\text{deg} \in [-21, -15]$, which contains approximately 10\thinspace{}000 stars.
We will do this using two data sets:
first, extinctions computed using the RJCE method; and second, extinctions reported by the StarHorse catalogue.
But before doing so we will test our method using synthetic data that mimic those generated by the RJCE method. 

To generate our synthetic data we realize the density field, $\density$, on a $501 \times 101 \times 45$ regular lattice in $s, l$, and $b$, under the assumption that $\density$ is a stationary and isotropic lognormal random process having a Kolmogorov-like power spectrum (Eq.~\ref{eq:psd_model}) with
mean
$\mu_{\density} = 1 \times 10^{-3}~\mathrm{mag}\thinspace\mathrm{pc}^{-1}$,
variance
$\sigma_{\densityfluctuations}^{2} = 1 \times 10^{-6}~\mathrm{mag}^{2}\mathrm{pc}^{-2}$,
energy-injection scale
$L_{0} = 100~\mathrm{pc}$, and
power-law indices
$\Omega = 1$ and
$\gamma = 11/3$.%
\footnote{The density field is then sampled at $1~\mathrm{pc}$ intervals in $s$ and $0.5~\mathrm{deg}$ intervals in $l$ and $b$.}
To do this, we note that the expectation and covariance of $\ln(\density(\bm{r}))$ are given by the well-known formulae \citep[see, for  example,][]{coles_nongaussian_1987}
\begin{align}
    \E{\ln(\density(\bm{r}))}
    = \ln(\E{\density(\bm{r})}) - \dfrac{1}{2}\var(\ln(\density(\bm{r})))
\end{align}
and
\begin{align}
    \cov(\ln(\density(\bm{r}_{i})), \ln(\density(\bm{r}_{j})))
    = \ln\left(\dfrac{\cov(\density(\bm{r}_{i}), \density(\bm{r}_{j}))}{\E{\density(\bm{r}_{i})}\E{\density(\bm{r}_{j})}} + 1\right)
\end{align}
and generate a realization of a Gaussian random process, $\ln(\density)$, which we then exponentiate, to give $\density$.
We then generate a realization of $A$ on the same $501 \times 101 \times 51$ lattice by approximating the line-of-sight integral using the trapezium rule, and choose a uniformly distributed random sample of this realization, size $n = 10~000$, to which we add normally distributed noise with zero mean and standard deviation of $0.1~\mathrm{mag}$.
This forms a set of of synthetic observations, $\bm{A} = (A_{1}, \dots , A_{n})$ at locations $\bm{r}_{1}, \dots , \bm{r}_{n}$ where we work in Galactic coordinates such that $\bm{r}_{i} = (s_{i}, l_{i}, b_{i})$.
To each distance we add normally distributed noise with zero mean and standard deviation of 5 per cent.
We then propagate errors from distance to extinction under the assumption that the linear approximation holds (Sec.~\ref{sec:computing_the_blup}).
Since the mean and maximum errors on distances are $19$~pc and $25$~pc respectively this approximation holds marginally.

\begin{figure}
  \centering
  \includegraphics[width=\columnwidth]{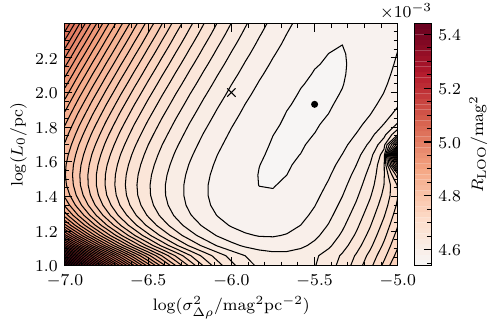}
  \caption{
  Recovery of the parameters of the autocovariance function for the case of synthetic data.
  The parameters of the autocovariance function, $\sigma^{2}_{\densityfluctuations}$ and $L_{0}$ (Eqs~\ref{eq:psd_model} and~\ref{eq:ext_covariance}), may be chosen so as to minimize the leave-one-out cross-validation score, $\score$ (Eq.~\ref{eq:loocv_score}).
  Here, a minimum in $\score$ is found for $\sigma^{2}_{\densityfluctuations} = \varsynthetic$ and $L_{0} = \scalesynthetic$.
  The true values of the parameters are $\sigma^{2}_{\densityfluctuations} = 1 \times 10^{-6}~\mathrm{mag}^{2}\mathrm{pc}^{-2}$ and $L_{0} = 100~\mathrm{pc}$.
  (See Sec.~\ref{sec:methodological_tests_using_synthetic_data} for discussion.)
}
  \label{fig:synthetic_data_parameter_estimation}
\end{figure}

\begin{figure}
  \centering
  \includegraphics[width=\columnwidth]{
    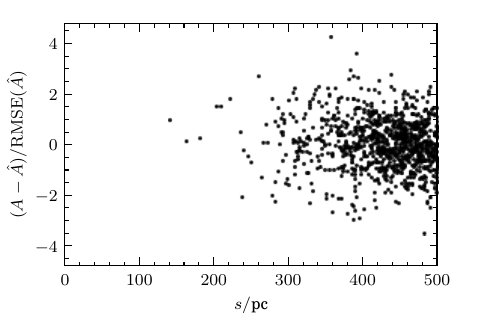
  }
  \includegraphics[width=\columnwidth]{
    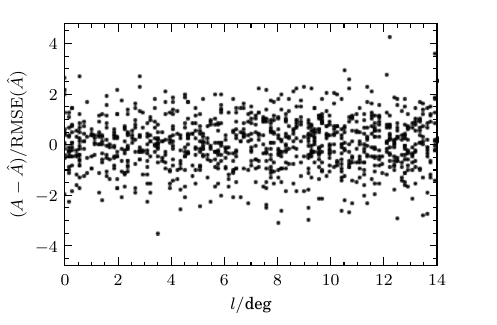
  }
  \includegraphics[width=\columnwidth]{
    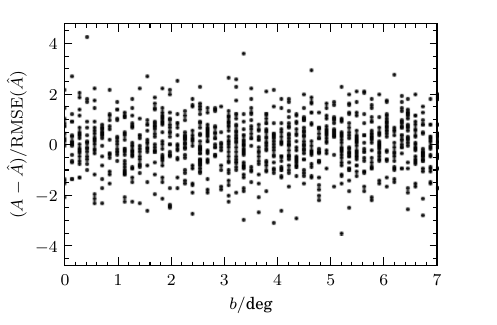
  }
  \caption{
  Validation of the best linear predictor for the case of synthetic data.
  The standardized leave-one-out residuals of the predicted extinction may be used to validate our maps.
  Here, these standardized residuals are seen to have approximately unit standard deviation, and exhibit no trends in $s$, $l$, or $b$.
  Our maps therefore pass validation.
  (For the sake of clarity these plots show a sample of size 1000, uniformly distributed in volume.)
  (See Sec.~\ref{sec:methodological_tests_using_synthetic_data} for discussion.)
  }
  \label{fig:synthetic_data_loocv}
\end{figure}

We now map the entire area by evaluating $\hat{A}$ and $\hat{\density}$ for all points on the lattice.
To do this, we assume that $\Omega = 1$ and $\gamma  = 11/3$, but optimize our choice of the remaining autocovariance parameters, $\sigma^{2}_{\densityfluctuations}$ and $L_{0}$, by minimizing the LOOCV score, $\score$ (Sec.~\ref{sec:choosing_the_model_parameter}).
This is expensive, since it involves the inversion of the matrix $\bm{\Sigma}$ for every trial parameter pair.
We therefore sample $\score$ on a $25 \times 25$ logarithmically spaced rectangular lattice covering the region $\sigma^{2}_{\densityfluctuations}/\mathrm{mag}^{2}\mathrm{pc}^{-2} \in [10^{-8}, 10^{-4}]$ and $L_{0}/\mathrm{pc} \in [10, 250]$.
Doing this, we find that $\sigma^{2}_{\densityfluctuations} = \varsynthetic$ and $L_{0} = \scalesynthetic$ (Fig.~\ref{fig:synthetic_data_parameter_estimation}).
These are close to the true values but we may not say they are consistent with the true values in any formal sense since we have no confidence region for them.
We then perform LOOCV (Section~\ref{sec:validation}) by inspecting the standardized leave-one-out residuals (Fig.~\ref{fig:synthetic_data_loocv}).
These have approximately unit standard deviation, and exhibit no trends in $s$, $l$, or $b$.
Accordingly, we say that our predictors pass validation.
Immediately, we find that the generalized least squares estimate of the density is $\hat{\mu}_{\density} = 1.2 \times 10^{-3}~\mathrm{mag}\thinspace\mathrm{pc}^{-1}$ with root mean-square error (RMSE) \rmsesynthetic{}~$\mathrm{mag}\thinspace\mathrm{pc}^{-1}$ (App.~\ref{sec:appendix_2}), consistent with the true value of $1 \times 10^{-3}~\mathrm{mag}\thinspace\mathrm{pc}^{-1}$.

\begin{figure*}
  \captionbox{
    Our synthetic extinction field, $A$ (left), and the field recovered from a noisy sample of it using our method, $\hat{A}$ (right), in the plane $b = 3.5~\mathrm{deg}$. (See Sec.~\ref{sec:methodological_tests_using_synthetic_data} for discussion.)
  \label{fig:synthetic_data_extinction}
  }
  [\columnwidth]
  {
    \includegraphics{
      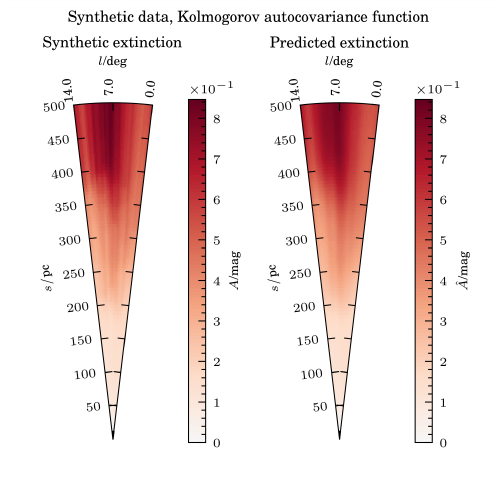
    }
  }
  \hfill
  \captionbox{
  Synthetic data:
  root mean square error, $\operatorname{RMSE}(\hat{A}) = \operatorname{std}(A - \hat{A})$, and standardized residuals of the predicted extinction for the plane $b = 3.5~\mathrm{deg}$. (See Sec.~\ref{sec:methodological_tests_using_synthetic_data} for discussion.)
  \label{fig:synthetic_data_extinction_errors}
  }
  [\columnwidth]
  {
    \includegraphics{
      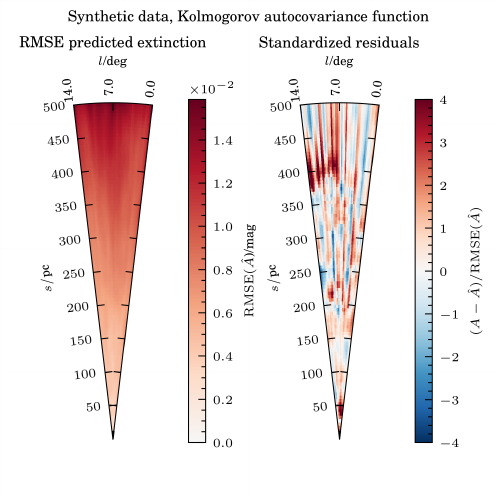
    }
  }
  \captionbox{
    Our synthetic density field, $\rho$ (left), and the field recovered from a sample of the extinction using our method, $\hat{\rho}$ (right), in the plane $b = 3.5~\mathrm{deg}$. (See Sec.~\ref{sec:methodological_tests_using_synthetic_data} for discussion.)
  \label{fig:synthetic_data_density}}
  [\columnwidth]
  {
    \includegraphics{
      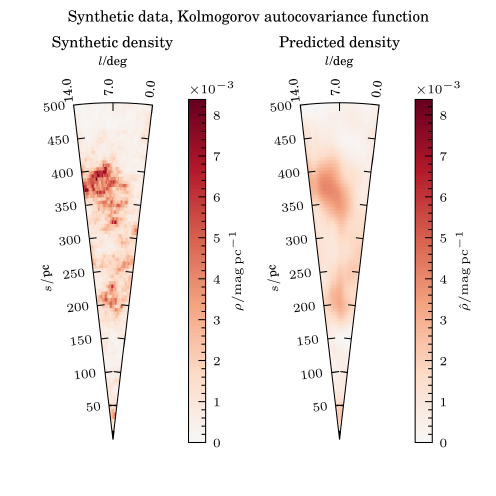
    }
  }
  \hfill
  \captionbox{
  Synthetic data:
  root mean square error, $\operatorname{RMSE}(\hat{\density}) = \operatorname{std}(\rho - \hat{\rho})$, and standardized residuals of the predicted density for the plane $b = 3.5~\mathrm{deg}$. (See Sec.~\ref{sec:methodological_tests_using_synthetic_data} for discussion.)
  \label{fig:synthetic_data_density_errors}
  }
  [\columnwidth]
  {
    \includegraphics{
      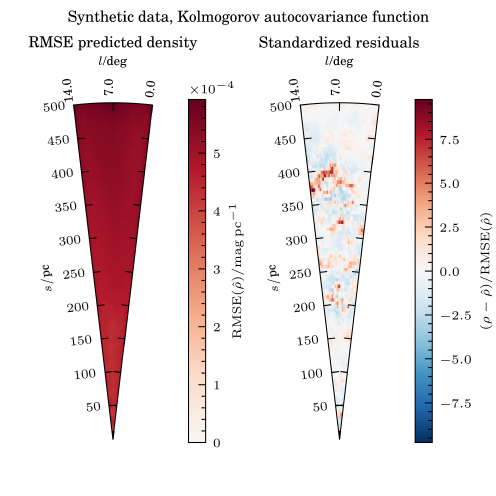
    }
  }
\end{figure*}

\begin{figure}
  \centering
  \includegraphics[width=\columnwidth]{
    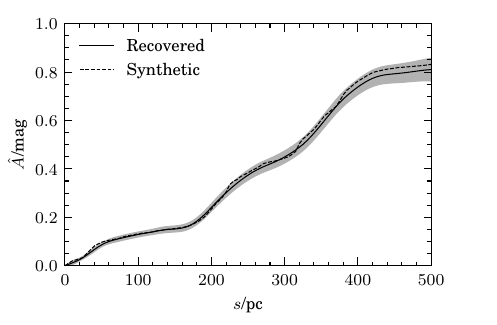
  }
  \includegraphics[width=\columnwidth]{
    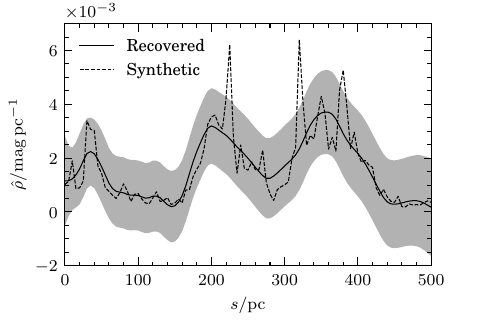
  }
  \caption{
  Our synthetic fields $A$ (top) and $\rho$ (bottom), along with the fields recovered using our method, along the line of sight $l = 7$~deg, $b = 3.5$~deg.
  In each case the synthetic field is shown as a dashed line, the recovered field is shown as a solid line, and the three-RMSE prediction interval shown as a shaded band. (See Sec.~\ref{sec:methodological_tests_using_synthetic_data} for discussion.)
  }
  \label{fig:synthetic_data_density_extinction_los}
\end{figure}

In Figures~\ref{fig:synthetic_data_extinction}--\ref{fig:synthetic_data_density_errors} we show our predictions for the plane $b = 3.5~\mathrm{deg}$, and in Figure~\ref{fig:synthetic_data_density_extinction_los} our predictions for the line of sight $l = 7~\mathrm{deg}$, $b = 3.5~\mathrm{deg}$.
Note that we cannot expect the BLUP to reproduce features on scales smaller than the mean stellar separation.
However, it is large fluctuations in the density on this interstellar scale that result in the largest increases in extinction.
These largest density fluctuations cannot be probed, and as a result the density map is smoothed out.
Nonetheless, the predictions for extinction and density are everywhere consistent with their true values.
Note that $\hat{A}$ is increasing in $s$, even though we do not constrain it to be.
Similarly, note that $\hat{\density}$ is non-negative for all $\bm{r}$.
These facts are most clearly seen in Figure~\ref{fig:synthetic_data_density_extinction_los}.
The prediction interval for extinction increases with distance and is inhomogeneous, just as the extinction is.
The prediction interval for density does not increase with distance and is homogeneous, just as the density itself is.
Note that predictions are even good close to the boundary, and in particular within the region for which $s < 200$~pc, where the predictor is most poorly constrained.

%% file: mapping_the_ism.tex
\label{sec:mapping_the_ism}

\newcommand{\nrjce}{\ensuremath{10\thinspace{}180}}
\newcommand{\nstarhorse}{\ensuremath{12\thinspace{}719}}
\newcommand{\nrjceneg}{\ensuremath{1\thinspace{}}434}
\newcommand{\nstarhorseneg}{\ensuremath{1\thinspace{}}434}
\newcommand{\varrjce}{\ensuremath{6.2 \times 10^{-5}~\mathrm{mag}^{2}\mathrm{pc}^{-2}}}
\newcommand{\varstarhorse}{\ensuremath{2.2 \times 10^{-5}~\mathrm{mag}^{2}\mathrm{pc}^{-2}}}
\newcommand{\scalerjce}{\ensuremath{98~\mathrm{pc}}}
\newcommand{\scalestarhorse}{\ensuremath{128~\mathrm{pc}}}
\newcommand{\rssrjce}{\ensuremath{0.012~\mathrm{mag}^{2}}}
\newcommand{\rssstarhorse}{\ensuremath{0.000~\mathrm{mag}^{2}}}
\newcommand{\bluerjce}{\ensuremath{3.8 \times 10^{-4}~\mathrm{mag}~\mathrm{pc}^{-1}}}
\newcommand{\bluestarhorse}{\ensuremath{6.2 \times 10^{-4}~\mathrm{mag}~\mathrm{pc}^{-1}}}
\newcommand{\bluecirjce}{\ensuremath{[-2.37 \times 10^{-3}, 3.13 \times 10^{-3}]}}
\newcommand{\rmserjce}{\ensuremath{2.75 \times 10^{-3}~\mathrm{mag}~\mathrm{pc}^{-1}}}
\newcommand{\bluecistarhorse}{\ensuremath{[-1.32 \times 10^{-3}, 2.56 \times 10^{-3}]}}
\newcommand{\rmsestarhorse}{\ensuremath{1.94 \times 10^{-3}~\mathrm{mag}~\mathrm{pc}^{-1}}}

Having tested our map-making method in this way we can now use it to create maps using each of our two data sets.

\begin{figure}
  \centering
  \includegraphics{
    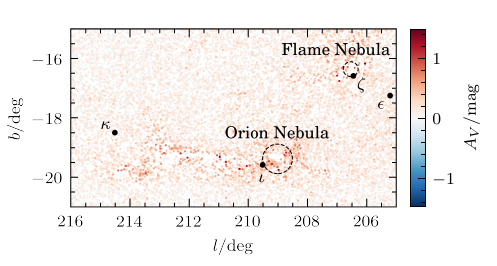
  }
  \caption{
    The $K_{\text{s}}$-band extinctions, computed using the RJCE method, for stars in the direction of the giant molecular cloud Orion A (note that the RJCE method permits unphysical negative extinctions).
  The principal stars of the Orion constellation ($\epsilon$~Ori, $\zeta$~Ori, $\kappa$~Ori, $\iota$~Ori) are marked.
  The head of Orion A is coincident with the Orion Nebula.
  Its tail extends south west (parallel to the Galactic plane) for several degrees, and is evident from the lack of observations in this region (due to high extinctions).
  The south-west extremity of the giant molecular cloud Orion B is coincident with the Flame Nebula.
  It extends north-west (perpendicular to the Galactic plane) for several degress, beyond the extent of the figure.
  It too is evident from a lack of observations.}
  \label{fig:rjce_extinctions_orion_a}
\end{figure}

\subsection{Application to colour-excess and \emph{Gaia} data}
\label{sec:mapping_the_ism_computed_extinctions}

For our RJCE data we use the catalogue compiled by \cite{rezaei_three-dimensional_2022}.%
\footnote{This catalogue is not publically available. We would like to thank Sara Rezaei Kh.\ for her generosity in making it available to us.}
This was generated using $H$-band observations from 2MASS \citep{skrutskie_two_2006} and $[4.5~\mu]$-band (i.e.\ W2-band) observations from WISE \citep{wright_wide-field_2010}, these catalogues 
having been cross-matched using the algorithm provided by the \gaia{} Archive.
The resulting sample was then cleaned of large-extinction outliers by eliminating all sources lying below the main-sequence in the dereddened colour-magnitude diagram \citep[see][for details]{rezaei_three-dimensional_2018}.
These large extinctions are either artefacts resulting from the combining of two catalogues, or are associated with young stellar objects and are hence due to dusty circumstellar discs and envelopes, not to the ISM itself.
We account for the errors in $H$ and $[4.5~\mu]$, which we denote $\sigma_{H}$ and $\sigma_{[4.5~\mu]}$, by adding them in quadrature, so that the variance in  $A_{K_{\mathrm{s}}}$ becomes $0.918^{2}\sigma^{2}_{H} + 0.918^{2}\sigma^{2}_{[4.5~\mu]} + 0.1^{2}~\mathrm{mag}^{2}$.
The resulting mean error in extinction is 0.17~mag, and the standard deviation of the error in extinction is 0.073~mag.

We combine these extinction data with the line-of-sight distances inferred by \cite{bailer-jones_estimating_2021} using \gaia{} EDR3 \citep{gaia_collaboration_gaia_2018}, again using the \gaia{} Archive cross-matching algorithm.
We take for our distance prediction the median of the posterior distribution (which we expect to be close to the mean of the predicted distance since the probability density functions are unimodal and approximately symmetric at such small distances), and for our error half the difference of the 0.16 and 0.84 quantiles.
The mean error in distance is 6.3~pc, and the standard deviation of the error in distance is 11~pc.
We propagate errors from distance to extinction under the assumption that the linear approximation holds (Sec.~\ref{sec:computing_the_blup}).
The final sample contains $\nrjce$ stars, having an average stellar separation of 3.9~pc.
Of these sources, $\nrjceneg$ have negative measured extinctions.
We show all extinctions in Fig.~\ref{fig:rjce_extinctions_orion_a}.

\begin{figure}
  \centering
  \includegraphics[width=\columnwidth]{
    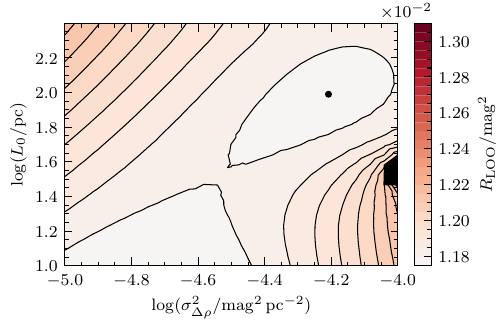
  }
  \caption{
  Recovery of the parameters of the autocovariance function for the case of RJCE and \gaia{} data.
  The leave-one-out cross-validation score, $\score$, is found to have a minimum for $\sigma^{2}_{\densityfluctuations} = \varrjce$ and $L_{0} = \scalerjce$.
  Contours are shown at intervals of $5 \times 10^{-5}~\mathrm{mag}^{2}$.
  (See Sec.~\ref{sec:mapping_the_ism} for discussion.)
  }
  \label{fig:rjce_data_parameter_estimation}
\end{figure}

\begin{figure*}
  \captionbox{
  Colour excess and \gaia{} data:
  predicted extinction, $\hat{A}_{K_{\mathrm{s}}}$ (left), and its root mean square errors, $\operatorname{RMSE}(\hat{A}_{K_{\mathrm{s}}})$ (right), for the plane $b = -19.5~\mathrm{deg}$.
  (See Sec.~\ref{sec:mapping_the_ism} for discussion.)
  \label{fig:rjce_data_extinction}
  }
  [\columnwidth]
  {
    \includegraphics{
      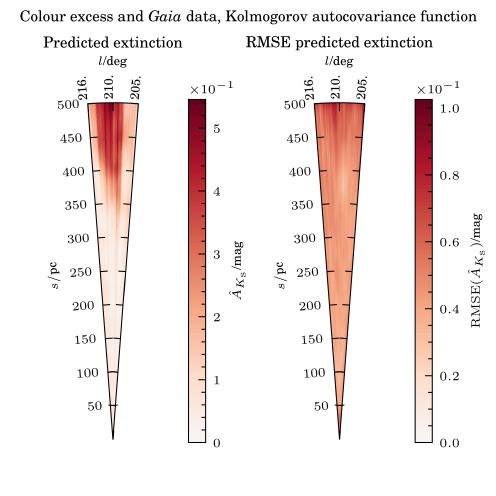
    }
  }
  \hfill
  \captionbox{
  Colour excess and \gaia{} data:
  predicted density, $\hat{\density}$ (left), and its root mean square errors, $\operatorname{RMSE}(\hat{\density})$ (right), for the plane $b = -19.5~\mathrm{deg}$.
  The head of Orion A is seen at $s = 390~\mathrm{pc}$ and a foreground cloud at $s = 350~\mathrm{pc}$.
  (See Sec.~\ref{sec:mapping_the_ism} for discussion.)
  \label{fig:rjce_data_density}
  }
  [\columnwidth]
  {
    \includegraphics{
      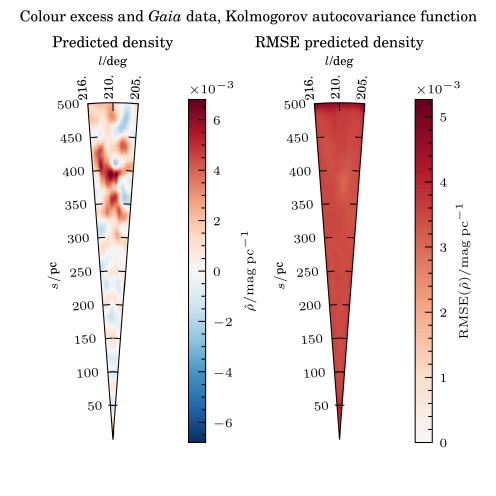 
    }
  }
\end{figure*}

\begin{figure*}
  \includegraphics{
    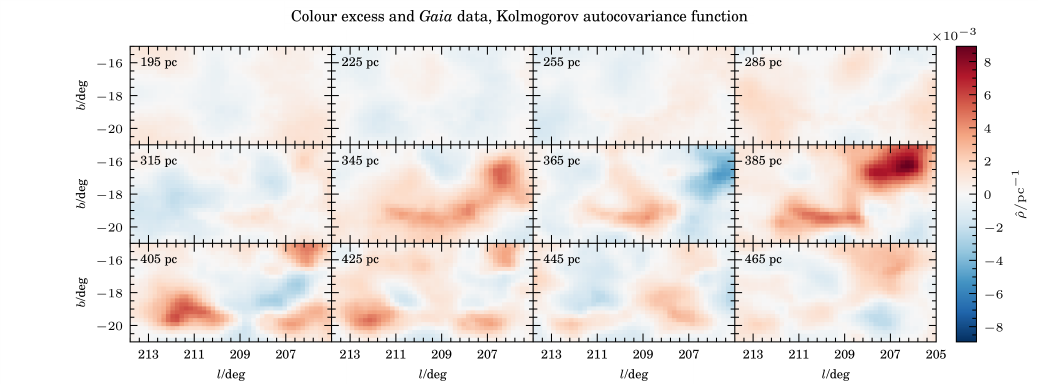
  }
  \caption{
  Colour excess and \gaia{} data:
  predicted density, $\hat{\density}$, for a series of line-of-sight distances, $s$, evaluated on a regular lattice with spacing of 0.5~deg.
  This figure matches Figure~2 in the paper by \protect{\citealt{rezaei_detailed_2020}}.
  The head of Orion A is seen in panels $s = 385~\mathrm{pc}$ and $s = 405~\mathrm{pc}$.
  A foreground cloud is seen in the panel $s = 345~\mathrm{pc}$.
  (See Sec.~\ref{sec:mapping_the_ism} for discussion.)
 }
  \label{fig:rjce_data_density_rezaei}
\end{figure*}

We again map the entire region by evaluating $\hat{A}$ and $\hat{\density}$ at all points on a $501 \times 501 \times 45$ regular lattice in $s$, $l$, and $b$, repeating the analysis we performed using synthetic data.
By minimizing the LOOCV score, $\score$, we find that that
$\sigma^{2}_{\densityfluctuations} = \varrjce$ and
$L_{0} = \scalerjce$
(Fig.~\ref{fig:rjce_data_parameter_estimation}).
(Note that the precision of these values is limited by the size of the grid we use to search parameter space.)
The leave-one-out residuals have better than unit standard deviation, and again exhibit no trends in $s$, $l$, or $b$, so that we may again say that our predictors pass validation (Fig.~\ref{fig:rjce_data_loocv}).
The GLS estimate of the density is $\hat{\mu}_{\density} = \bluerjce$ with RMSE \rmserjce (see App.~\ref{sec:appendix_2} for a discussion of our quoted errors).
In Figures~\ref{fig:rjce_data_extinction} and \ref{fig:rjce_data_density} we show our predictions for the plane $b = -19.5~\mathrm{deg}$ (corresponding to Fig.~4 in the paper by \citealt{grosschedl_vision_2019}), and in Figure~\ref{fig:rjce_data_density_rezaei} we show our predictions for a sequence of on-sky regions at increasing line-of-sight distances (corresponding to Figs~2 and~5 in the 2020 paper by \citeauthor{rezaei_detailed_2020}).
Note that $\hat{A}$ is not increasing in $s$, and that $\hat{\density}$ is negative for some $\bm{r}$.
Nevertheless, prediction intervals for $\hat{A}$ are consistent with its being increasing in $s$, and prediction intervals for $\hat{\density}$ are consistent with its being non-negative for all $\bm{r}$.

\begin{figure}
  \centering
  \includegraphics[width=\columnwidth]{
    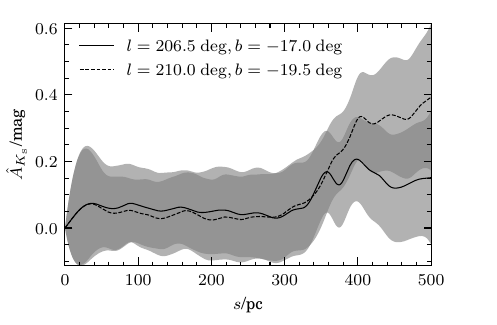
  }
  \includegraphics[width=\columnwidth]{
    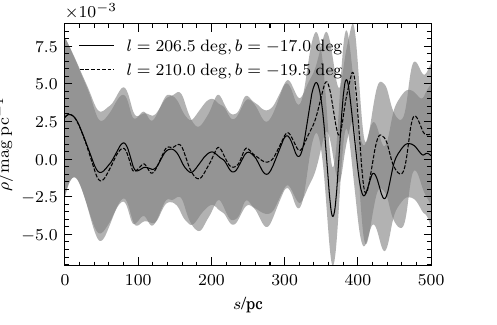
  }
  \caption{
  Colour excess and \gaia{} data:
  extinction, $A_{K_{\mathrm{s}}}$ (top), density, $\density$ (bottom), and their predicted values, $\hat{A}_{K_{\mathrm{s}}}$ and $\hat{\density}$, for the lines of sight $l = 206.5~\mathrm{deg}$, $b = -17~\mathrm{deg}$ and $l = 210~\mathrm{deg}$, $b = -19.5~\mathrm{deg}$.
  These pass through the two lobes of the foreground cloud.
  The three-RMSE prediction intervals are shown as shaded bands.
  (See Sec.~\ref{sec:mapping_the_ism} for discussion.)
  }
  \label{fig:rjce_data_density_extinction_los}
\end{figure}

Our maps clearly identify the head of Orion A (at $s = 390~\mathrm{pc}$, $l = -19.5~\mathrm{deg}$, $b = 209~\mathrm{deg}$) and its tail, which extends to a distance of $s = 470~\mathrm{pc}$ albeit varying in density along its length.
They also clearly identify the foreground cloud first noted by \cite{rezaei_detailed_2020}.
In the plane $b = -19.5~\mathrm{deg}$ (Fig.~\ref{fig:rjce_data_density}) this appears as a chevron, the apex of which is located at $s = 350~\mathrm{pc}$, $l = 210~\mathrm{deg}$.
In the plane $s = 345~\mathrm{pc}$ (Fig.~\ref{fig:rjce_data_density_rezaei}) it is seen to have two lobes, one centred at $l = -19.5~\mathrm{deg}$, $b = 211~\mathrm{deg}$, and the other centred at $l = 206.5~\mathrm{deg}$, $b = 17~\mathrm{deg}$.
The lower of these is located at the apex of the chevron, and is bifurcated at smaller distances.
Figure~\ref{fig:rjce_data_density_extinction_los} shows our predictions for the lines of sight centred on these lobes (corresponding to Fig.~3 in the 2020 paper by \citeauthor{rezaei_detailed_2020}).

Let us consider possible systematic errors in our predictions.
These might arise from inadequate models of the mean and covariance of the extinction field (Eqs~\ref{eq:mean_density} and \ref{eq:cov_ext}) or from systematic errors in the distance and extinction data that we use.
Such errors might be associated with stars of a particular spectral class, causing them to appear too close or too distant, too red or too blue. 
The distances to our sources have been inferred using only parallax information so we do not expect there to be any systematic errors of this type in distance.
However, the RJCE method for computing extinctions relies on a linear fit to observed colours.
This assumption of linearity is good for a large range of spectral classes but could result in the extinctions of very cool stars being underestimated and the extinctions of very hot stars being be overestimated.
Sytematic errors associated with spectral class should appear as overdensities in the joint distribution of (i) spectral class and distance or (ii) spectral class and extinction, and such overdensities should alert us to the presence of systematic errors in the computed values of either quantity.
We can use the dereddened colour $(G - K_{\text{s}})_{0}$ (computed using using the exinction law of \citealt{cardelli_relationship_1989}) as a proxy for spectral class and inspect the histogram for each distribution (not shown).
In neither case do we see overdensities associated with any particular value of colour. We therefore claim that our data are free of significant systematic errors relating to spectral class.

\subsection{Application to StarHorse data}
\label{sec:mapping_the_ism_inferred_extinctions}

The StarHorse catalogue includes quantiles for the inferred probability density functions of a source's extinction and distance, and therefore neatly provides all the information we need to construct a dust map.
We use the first StarHorse catalogue \citep{anders_photo-astrometric_2019} rather than the second \citep{anders_photo-astrometric_2021} since this second catalogue begs the question by assuming for a prior on extinction the maps of \citeauthor{green_3d_2019} and \citeauthor{drimmel_three-dimensional_2003}
Each source in the StarHorse catalogue is associated with the compound flags \texttt{SH\_GAIAFLAG}, which quantifies the quality of the \gaia{} astrometry and photometry used, and \texttt{SH\_OUTFLAG}, which quantifies the quality of StarHorse's inferences \citep{anders_photo-astrometric_2019}.
The \texttt{SH\_GAIAFLAG} flag consists of three digits flagging
(1) high renormalized unit weight error,
warning of a spurious astrometric solution,
(2) high colour excess factor,
warning of spurious photometry, and
(3) source variability.
The \texttt{SH\_OUTFLAG} flag consists of five digits flagging
(1) overall unreliability,
warning of physically inconsistent inferred stellar properties,
(2) large distance,
warning of a spuriously large inferred distance,
(3) unreliable extinction,
warning of a spuriously small or spuriously large inferred extinction,
(4) large extinction uncertainty,
warning of an inferred extinction uncertainty of unit magnitude or greater, and
(5) 
small uncertainty,
warning of spuriously small uncertainties in any inferred stellar property.
We use sources for which \texttt{SH\_GAIAFLAG} is \texttt{000} and \texttt{SH\_OUTFLAG} is \texttt{00000}, ensuring that potentially spurious statistics of each kind are not included in our sample.

For each source in this sample, we realize a sample of the posterior extinction by assuming its PDF to be uniform within quantiles.
In order to do this, we assume a lower limit on $A_{V}$ of $-0.3~\mathrm{mag}$, and an upper limit of $4~\mathrm{mag}$, consistent with the StarHorse prior PDF for extinctions with errors in distance greater that 20~per cent.
We take the standard deviation of the posterior extinction to be half the difference of the 0.16 and 0.84 quantiles.
The mean error in extinction is then 0.17~mag, and the standard deviation of the error in extinction is 0.011~mag.

We again take for our distance prediction the median of the posterior distribution, and for our error half the difference of the 0.16 and 0.84 quantiles.
The mean error in distance is then 13~pc, and the standard deviation of the error in distance is 9.7~pc.
Again, we propagate errors from distance to extinction under the assumption that the linear approximation holds (Sec.~\ref{sec:computing_the_blup}).
The final sample contains $\nstarhorse$ stars, having an average stellar separation of approximately 3.7~pc.
Of these sources, $\nstarhorseneg$ have negative extinctions.

\begin{figure}
  \centering
  \includegraphics[width=\columnwidth]{
    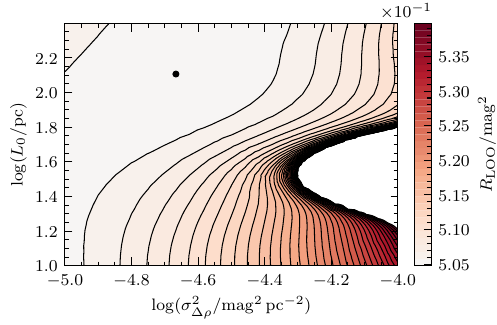
  }
  \caption{
  Recovery of the parameters of the autocovariance function for the case of StarHorse data.
  The leave-one-out cross-validation score, $\score$, is found to have a minimum for $\sigma^{2}_{\densityfluctuations} = \varstarhorse$ and $L_{0} = \scalestarhorse$.
  The covariance matrix, $\bm{\Sigma}$, is nearly singular for some parameter combinations (region shown in white), meaning that $\score$ cannot be computed.
  Contours are shown at intervals of $1 \times 10^{-3}~\mathrm{mag}^{2}$.
  (See Sec.~\ref{sec:mapping_the_ism} for discussion.)
  }
  \label{fig:starhorse_data_parameter_estimation}
\end{figure}

\begin{figure*}
  \captionbox{
  StarHorse data:
  predicted extinction, $\hat{A}_{V}$ (left), and its root mean square errors, $\operatorname{RMSE}(\hat{A}_{V})$ (right), for the plane $b = -19.5~\mathrm{deg}$.
  Cp.\ Fig.~\ref{fig:starhorse_data_extinction}.
  (See Sec.~\ref{sec:mapping_the_ism} for discussion.)
  \label{fig:starhorse_data_extinction}
  }
  [\columnwidth]
  {
    \includegraphics{
      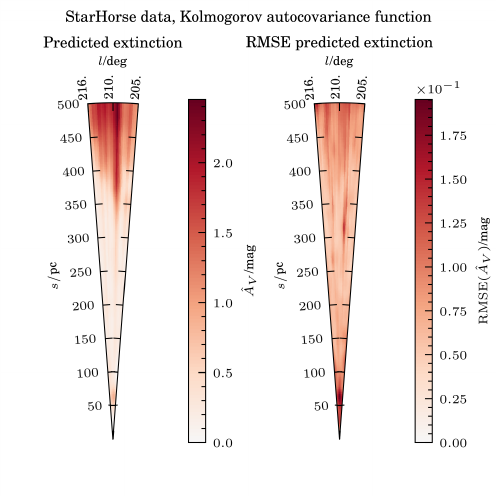
    }
  }
  \hfill
  \captionbox{
  StarHorse data:
  predicted density, $\hat{\density}$ (left), and its root mean square errors, $\operatorname{RMSE}(\hat{\density})$ (right), for the plane $b = -19.5~\mathrm{deg}$.
  Cp.\ Fig.~\ref{fig:rjce_data_density}.
  (See Sec.~\ref{sec:mapping_the_ism} for discussion.)
  \label{fig:starhorse_data_density}
  }
  [\columnwidth]
  {
    \includegraphics{
      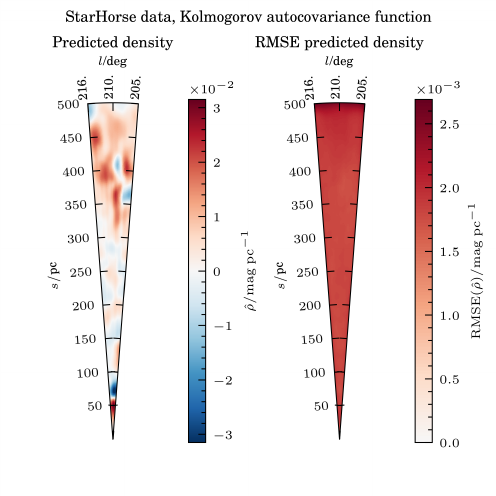
    }
  }
\end{figure*}

\begin{figure*}
  \includegraphics{
    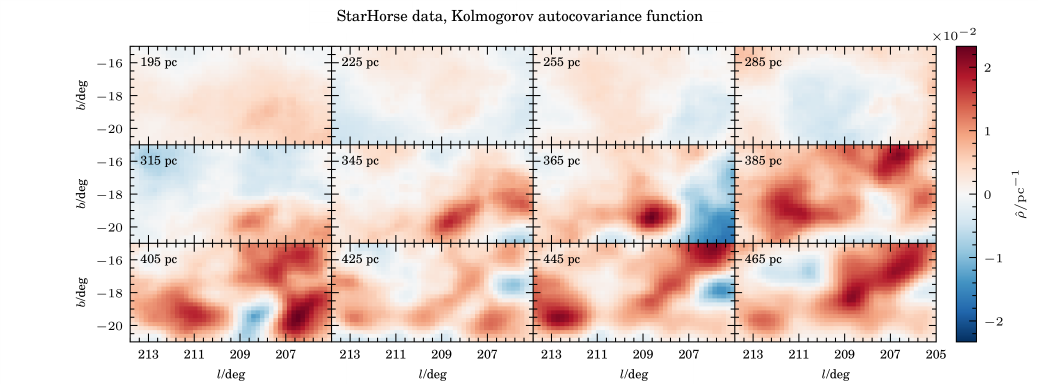
  }
  \caption{
  StarHorse data:
  predicted density, $\hat{\density}$, for a series of line-of-sight distances, $s$.
  Cp.\ Fig.~\ref{fig:rjce_data_density_rezaei}.
  (See Sec.~\ref{sec:mapping_the_ism} for discussion.)
  }
  \label{fig:starhorse_data_density_rezaei}
\end{figure*}

\begin{figure}
  \centering
  \includegraphics[width=\columnwidth]{
    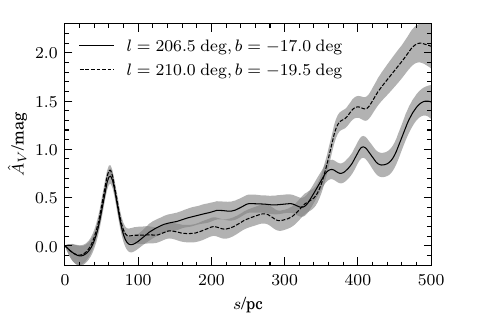
  }
  \includegraphics[width=\columnwidth]{
    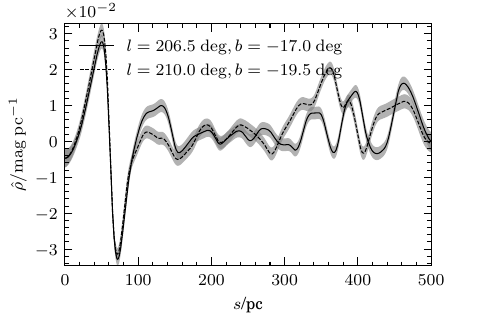
  }
  \caption{
  StarHorse data:
  extinction, $A_{V}$ (top), density, $\density$ (bottom), and their predicted values, $\hat{A}_{V}$ and $\hat{\density}$, for the lines of sight $l = 206.5~\mathrm{deg}$, $b = -17~\mathrm{deg}$ and $l = 210~\mathrm{deg}$, $b = -19.5~\mathrm{deg}$.
  The three-RMSE prediction intervals are shown as shaded bands.
  Cp.\ Fig.~\ref{fig:rjce_data_density_extinction_los}.
  (See Sec.~\ref{sec:mapping_the_ism} for discussion.)
  }
  \label{fig:starhorse_data_density_los}
\end{figure}

By minimizing the LOOCV, $\score$, we find that that $\sigma^{2}_{\densityfluctuations} = \varstarhorse$, and $L_{0} = \scalestarhorse$ (Fig.~\ref{fig:starhorse_data_parameter_estimation}).
However, the leave-one-out residuals have very high standard deviation of $4.5~\mathrm{mag}^{2}$, although they exhibit no trends in $s$, $l$, or $b$ (Fig.~\ref{fig:starhorse_data_loocv}).
As such, they fail validation.
This may be due to the presence of young stellar objects in the StarHorse catalogue, which we have made no attempt to identify and remove, and which violate the assumptions of our model.
Nevertheless, we find that GLS estimate of the density is $\hat{\mu}_{\density} = \bluestarhorse$ with RMSE \rmsestarhorse{} and, as before, we plot our predictions for the plane $b = -19.5~\mathrm{deg}$ (Figs~\ref{fig:starhorse_data_extinction} and~\ref{fig:starhorse_data_density}), a sequence of on-sky regions at increasing line-of-sight distances (Fig.~\ref{fig:starhorse_data_density_rezaei}), and the lines of sight $l = 206~\mathrm{deg}$, $b = -17~\mathrm{deg}$, and $l = 210~\mathrm{deg}$, $b = -19.5~\mathrm{deg}$ (Fig.~\ref{fig:starhorse_data_density_los}).
Note that StarHorse extinctions are given for the $V$-band rather than $K_{\mathrm{s}}$-band, and that according to the \cite{cardelli_relationship_1989} extinction law, $A_{V}/A_{K_{\mathrm{s}}} = 8.8$.
Again, we note that $\hat{A}$ is not increasing in $s$, and $\hat{\density}$ is negative for some $\bm{r}$.
Nevertheless, prediction intervals for $\hat{A}$ are again consistent with its being increasing in $s$, and prediction intervals for $\hat{\density}$ are again consistent with its being non-negative for all $\bm{r}$.

The head of Orion A (at $s = 390~\mathrm{pc}$, $l = 209~\mathrm{deg}$, $b = -19.5~\mathrm{deg}$) appears only faintly in our maps, although the tail and foreground cloud both appear clearly.
In the plane $b = -19.5~\mathrm{deg}$ the foreground cloud is seen as a linear structure rather than chevron, though it is again seen to have two lobes in the plane $s = 345~\mathrm{pc}$.
Since these maps do not pass validation we do not use them to make comment on the distribution of dust in Orion A. In particular we do not use them to comment on the existence of a foreground cloud.

Here we have used a single realization of the StarHorse posterior extinctions.
Alternative realizations of the posterior extinctions produce similar results.
We could instead have made multiple realizations and generated an empirical distribution for the predictor $\hat{A}(\bm{r})$.
In practice, however, we find that the mean and standard deviation of the empirical distribution converge quickly, and that they are in fact well approximated by a single sample.

%% file: discussion.tex
\label{sec:discussion}

\newcommand{\varrjcegneitinga}{\ensuremath{1.78 \times 10^{-5}}~\mathrm{mag}^{2}}
\newcommand{\varrjcegneitingb}{\ensuremath{5.62 \times 10^{-5}}~\mathrm{mag}^{2}}
\newcommand{\scalerjcegneitinga}{\ensuremath{13.1}~\mathrm{pc}}
\newcommand{\scalerjcegneitingb}{\ensuremath{191}~\mathrm{pc}}
\newcommand{\varrjcese}{\ensuremath{1 \times 10^{-5}~\mathrm{mag}^{2}}}
\newcommand{\scalerjcese}{\ensuremath{40~\mathrm{pc}}}

To make our maps we have used a physically motivated Kolmogorov-like autocovariance function to describe the density fluctuations of the ISM.
When using the BLUP (Eq.~\ref{eq:blup}) it is common to use one of many off-the-shelf library functions that do not have direct physical motivation.
For example, it is very common to use a powered-exponential autocovariance function,
$\autocovprime{\mathrm{PE}}$,
given by 
\begin{align}
  \label{eq:powered_exponential_function}
  \autocovprime{\mathrm{PE}}(\delta) = \sigma^{2}_{\mathrm{PE}}\exp\left(-\left(\dfrac{\delta}{L_{\mathrm{PE}}}\right)^{\alpha}\right)
\end{align}
where $\sigma^{2}_{\mathrm{PE}}$ is the variance of the field, $L_{\mathrm{PE}}$ is a characteristic length scale, and where $\alpha \in (0, 2]$.
Particularly popular is the case of $\alpha = 2$, when $\autocovprime{\mathrm{PE}}$ is known as the `squared-exponential autocovariance function'.
Autocovariance functions with compact support are also attractive, since they reduce the computational burden of computing and inverting the covariance matrix, $\bm{\Sigma}$.
Indeed \cite{rezaei_inferring_2017} recommended the use of compactly supported functions for dust mapping, and in their work use Gneiting's function, $\autocovprime{\mathrm{G}}$ \citep{gneiting_compactly_2002}, given by
\begin{align}
  \label{eq:gneiting_function}
  \autocovprime{\mathrm{G}}(\delta) = 
  \sigma^{2}_{\mathrm{G}}
  \left(1 + \left(\dfrac{\delta}{L_{\mathrm{G}}}\right)\right)^{-3}
  \left(\left(1 - \dfrac{\delta}{L_{\mathrm{G}}}\right)\cos\left(\dfrac{\pi{}\delta}{L_{\mathrm{G}}}\right) + \dfrac{1}{\pi}\sin\left(\dfrac{\pi{}\delta}{L_{\mathrm{G}}}\right)\right)
\end{align}
if $\delta/L_{\mathrm{G}} \in [0, 1]$, and $\autocovprime{\mathrm{G}}(\delta) = 0$ otherwise, where $\sigma^{2}_{\mathrm{G}}$ is the variance of the field and $L_{\mathrm{G}}$ is a characteristic length scale for the field, above which the autocovariance vanishes.

Any such standard autocovariance function may be used with our method by integrating it to give an approximation to $\cov(A(\bm{r}_{i}), A(\bm{r}_{j}))$ according to Equation~\ref{eq:ext_covariance}.
In turn, we can approximate the functions $f$ and $g$ (Eqs~\ref{eq:function_f} and~\ref{eq:function_g}).
(See App.~\ref{sec:appendix_2} for plots of these approximations.)
Although library functions of this kind have no direct physical motivation, we do require them to have properties consistent with the physics of the ISM.
In particular we require the random processes they define to be continuous, undifferentiable and to have physically meaningful characteristic length scales.
We should be mindful of these properties when we specify the autocovariance function and when we interpret our predictions.
A stationary and isotropic random field is continuous (in the mean square) if its autocovariance function is continuous at zero,
and is differentiable (in the mean square) if its autocovariance function is also twice differentiable at zero (i.e.\ a stationary random field is $m$-times differentiable everywhere if its autocovariance function is $2m$-times differentiable at zero).
The Kolmogorov-like autocovariance function is undifferentiable at zero, and hence defines an undifferentiable random field.
This roughness is clear in our synthetic dust maps (bottom panel of Fig.~\ref{fig:synthetic_data_density_extinction_los}).
Gneiting's function is also undifferentiable at zero, and therefore also defines an undifferentiable random field.
The powered-exponential is differentiable only for $\alpha = 2$, when it is infinitely differentiable, and hence defines an infinitely differentiable random field.
For $\alpha \neq 2$, however, it is undifferentiable, and hence again defines an undifferentiable random field.
Indeed, realizations of the random field increase in roughness as $\alpha$ decreases.
Whereas Gneiting's function is suitable for specifying the dust's autocovariance, the squared-exponential kernel is not.
Using it will result in maps that lack sharpness and exhibit features that are ill-defined.

To find a physically meaningful interpretation of the characteristic length scale we may use the \emph{integral length scale} of a random field, which is much used in the study of turbulence \citep{tennekes_first_1972}.
Suppose that $\autocovprime{}$ is a autocovariance function with variance $\sigma^{2}$ and characteristic length scale $L$.
The integral length scale of the stationary random field associated with $\autocovprime{}$ is%
\footnote{Note that this integral always exists since $\autocovprime{}$ is absolutely integrable.}
\begin{align}
  l(\autocovprime{}) = \dfrac{1}{\sigma^{2}}\int_{0}^{\infty}|\autocovprime{}(\delta)|\diff{}\delta.
\end{align}
In order to make explicit its dependence on $\sigma^{2}$ and $L$ let us write $\autocovprime{}$ as $\autocovprime{}(\cdot; \sigma^{2}, L)$.
We introduce the \emph{physical length scale}, $\Lambda$, such that 
\begin{align}
  \dfrac{1}{\sigma^{2}}\int_{0}^{\infty}|\autocovprime{}(\delta; \sigma^{2}, L)|\diff{}\delta
  = \dfrac{1}{\sigma^{2}_{\densityfluctuations}}\int_{0}^{\infty}|\autocovprime{\densityfluctuations}(\delta; \sigma^{2}_{\densityfluctuations}, \Lambda)|\diff{}\delta,
\end{align}
noting that, whereas $k$ is the autocovariance function of interest, $k_{\densityfluctuations}$ is the autocovariance of the density fluctuations (eq.~\ref{eq:autocorrelation_model}) with power spectrum defined by \cite{sale_three-dimensional_2014} and given in equation~\ref{eq:psd_model}.
This may be rewritten as
\begin{align}
  \dfrac{L}{\sigma^{2}}\int_{0}^{\infty}|\autocovprime{}(\delta; \sigma^{2}, 1)|\diff{}\delta
  = \dfrac{\Lambda}{\sigma^{2}_{\densityfluctuations}}\int_{0}^{\infty}|\autocovprime{\densityfluctuations}(\delta; \sigma^{2}_{\densityfluctuations}, 1)|\diff{}\delta,
\end{align}
or, equivalently,
\begin{align}
  \label{eq:physical_length_scale}
  \Lambda = \dfrac{\int_{0}^{\infty}|\autocovprime{}(\delta; 1, 1)|\diff{}\delta}{\int_{0}^{\infty}|\autocovprime{\densityfluctuations}(\delta; 1, 1)|\diff{}\delta}L,
\end{align}
which allows us to find the physical length scale associated with a given autocovariance function.
It approximates the energy-injection scale of the equivalent Kolmogorov-like density field.
The Kolmogorov-like, squared-exponential, and Gneiting functions have physical length scales
$L_{0}$,
$2.51L_{\mathrm{PE}}$, and
$0.466L_{\mathrm{G}}$.

\begin{figure}
  \centering
  \includegraphics[width=\columnwidth]{
    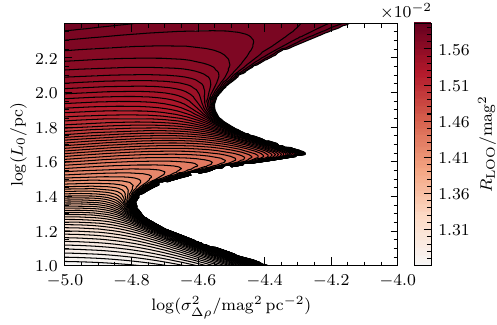
  }
  \includegraphics[width=\columnwidth]{
    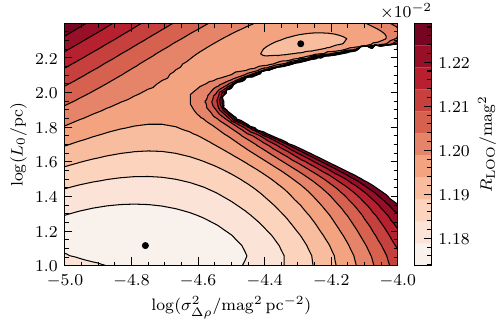
  }
  \caption{
    Top: recovery of the parameters of squared-exponential autocovariance function (Eq.~\ref{eq:powered_exponential_function}, $\alpha = 2$).
  The leave-one-out cross-validation score, $\score$, is found to have no minimum.
  Bottom: recovery of the parameters of Gneiting's autocovariance function (Eq.~\ref{eq:gneiting_function}).
  The leave-one-out cross-validation score, $\score$, is found to have a minumum for $\sigma^{2}_{\densityfluctuations} = \varrjcegneitinga$, $L_{0} = \scalerjcegneitinga$ and $\sigma^{2}_{\densityfluctuations} = \varrjcegneitingb$, $L_{0} = \scalerjcegneitingb$.
  In both cases, the covariance matrix, $\bm{\Sigma}$, is nearly singular for some parameter combinations (regions shown in white), meaning that $\score$ cannot be computed.
  (Cp. Fig.~\ref{fig:rjce_data_parameter_estimation}.)
  Contours are shown at intervals of $5 \times 10^{-5}~\mathrm{mag}^{2}$.
  (See Sec.~\ref{sec:discussion} for discussion.)
  }
  \label{fig:rjce_data_parameter_estimation_gneiting_se}
\end{figure}

\begin{figure*}
  \captionbox{
  Colour excess and \gaia{} data:
  predicted extinctions, $\hat{A}_{K_{\mathrm{s}}}$, and their root mean square errors, $\operatorname{RMSE}(\hat{A}_{K_{\mathrm{s}}})$, computed using the squared-exponential function (Eq.~\ref{eq:powered_exponential_function}), for the plane $b = -19.5~\mathrm{deg}$.
  (See Sec.~\ref{sec:discussion} for discussion. Cp. Fig.~\ref{fig:rjce_data_extinction}, which shows predictions computed using the Kolmogorov-like function.)
  \label{fig:rjce_data_extinction_se}
  }
  [\columnwidth]
  {
    \includegraphics{
      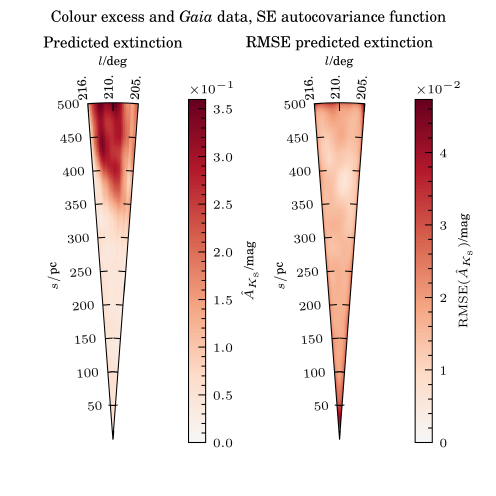
    }
  }
  \hfill
  \captionbox{
  Colour excess and \gaia{} data:
  predicted density, $\hat{\density}$, and their root mean square errors, $\operatorname{RMSE}(\hat{\density})$, computed using the squared-exponential function (Eq.~\ref{eq:powered_exponential_function}), for the plane $b = -19.5~\mathrm{deg}$.
  (See Sec.~\ref{sec:discussion} for discussion. Cp. Fig.~\ref{fig:rjce_data_density}, which shows predictions computed using the Kolmogorov-like function.) 
  \label{fig:rjce_data_density_se}
  }
  [\columnwidth]
  {
    \includegraphics{
      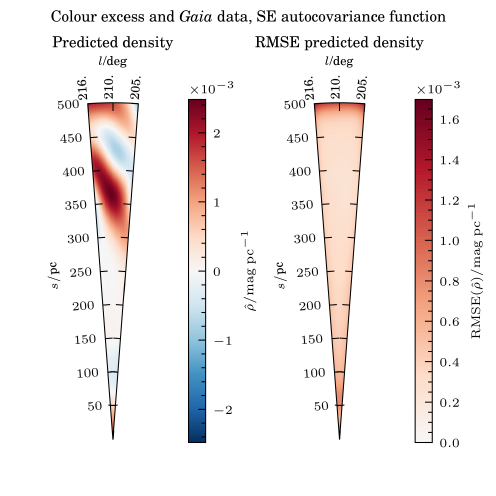
    }
  }
\end{figure*}

We can illustrate the consequences of our choice of autocovariance function by making maps using the squared-exponential and Gneiting functions with the data of \cite{rezaei_three-dimensional_2022} and \cite{bailer-jones_estimating_2021} (Sec.~\ref{sec:mapping_the_ism}).
For the squared-exponential function we find no minimum in the LOOCV score:
for large regions of parameter space inversion of $\bm{\Sigma}$ is numerically unstable (and hence the LOOCV may not be computed),
whereas elsewhere the LOOCV score reduces monotonically as $L_{\mathrm{PE}}$ approaches $10~\mathrm{pc}$ (the lower limit of the search region) for all values of $\sigma^{2}_{\mathrm{PE}}$ (Fig.~\ref{fig:rjce_data_parameter_estimation_gneiting_se}, top panel).
This reflects the fact that the autocovariance function is misspecified, and that the LOOCV score of the BLUP can only be reduced by its overfitting the data.
We may nonetheless choose the parameters of our function arbitrarily.
Suppose, then, that we choose $\sigma^{2}_{\mathrm{PE}} = \varrjcese$ and $L_{\mathrm{PE}} = \scalerjcese$, such that $L_{\mathrm{PE}}$ is the characteristic length we would expect if $\Lambda = 100~\mathrm{pc}$.
In this case our predictors do pass validation, and we are able to make the maps shown in Figures~\ref{fig:rjce_data_extinction_se} and~\ref{fig:rjce_data_density_se}.
Note that the features in our map (in particular, the head and tail of Orion A, as well as the foreground cloud) are no longer resolved and that the map itself is much less sharp than before, reflecting the fact that the squared-exponential function results in smooth random fields that do not have the rough properties of the ISM, and that because of this its length scale cannot be given a physical interpretation.
Moreover, our results are sensitive to our choice of length scale, meaning that a scaling of $L_{\mathrm{PE}}$ will result in a scaling of the predicted features.
With a characteristic length scale of $L_{\mathrm{PE}} = 15~\text{pc}$ our predictors again pass validation, and the features in our map are resolved, but still lack sharpness.
(We do not show this effect in our figures.)

\begin{figure*}
  \captionbox{
  Colour excess and \gaia{} data:
  predicted extinctions, $\hat{A}_{K_{\mathrm{s}}}$, and their root mean square errors, $\operatorname{RMSE}(\hat{A}_{K_{\mathrm{s}}})$, computed using the Gneiting function (Eq.~\ref{eq:gneiting_function}), for the plane $b = -19.5~\mathrm{deg}$.
  (See Sec.~\ref{sec:discussion} for discussion. Cp. Fig.~\ref{fig:rjce_data_extinction}, which shows predictions computed using the Kolmogorov-like function.)
  \label{fig:rjce_data_extinction_gneiting}
  }
  [\columnwidth]
  {
    \includegraphics{
      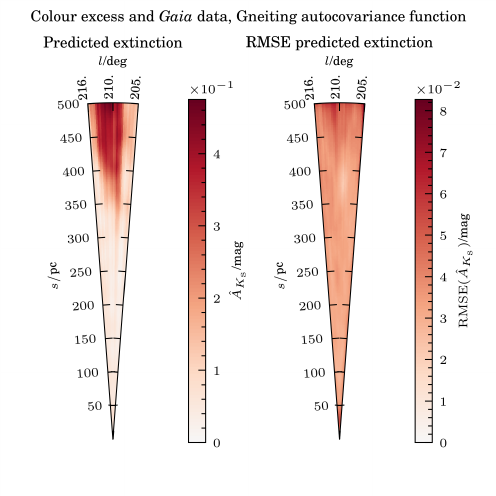
    }
  }
  \hfill
  \captionbox{
  Colour excess and \gaia{} data:
  predicted density, $\hat{\density}$, and their root mean square errors, $\operatorname{RMSE}(\hat{\density})$, computed using the Gneiting function (Eq.~\ref{eq:gneiting_function}), for the plane $b = -19.5~\mathrm{deg}$.
  (See Sec.~\ref{sec:discussion} for discussion. Cp. Fig.~\ref{fig:rjce_data_density}, which shows predictions computed using the Kolmogorov-like function.)
  \label{fig:rjce_data_density_gneiting}
  }
  [\columnwidth]
  {
    \includegraphics{
      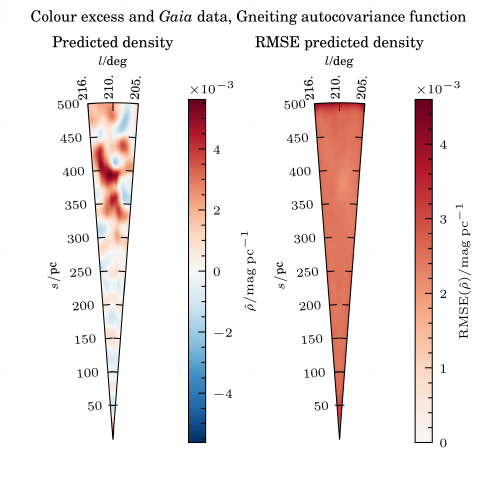
    }
  }
\end{figure*}

For Gneiting's function we find two minima in the LOOCV score, at $\sigma^{2}_{\mathrm{G}} = \varrjcegneitinga$, $L_{\mathrm{G}} = \scalerjcegneitinga$, and $\sigma^{2}_{\mathrm{G}} = \varrjcegneitingb$, $L_{\mathrm{G}} = \scalerjcegneitingb$ (Fig.~\ref{fig:rjce_data_parameter_estimation_gneiting_se}, bottom panel).
Both pairs of parameters result in predictors that pass validation.
However, it is the second pair that corresponds to the physical length scale we expect, since in this case we find that $\Lambda = 89~\mathrm{pc}$.
Maps made using this second pair of parameters are shown in Figures~\ref{fig:rjce_data_extinction_gneiting} and~\ref{fig:rjce_data_density_gneiting}.
They are all but indistinguishable from maps made using the first pair of parameters, indicating that our predictors are somewhat robust against misspecification of the length scale.
Moreover, they are very similar to the maps made using the Kolmogorov-like autocovariance function (Figs~\ref{fig:rjce_data_extinction} and~\ref{fig:rjce_data_density}), indicating that Gneiting's function provides a good approximation to the Kolmogorov-like function.

%% file: conclusion.tex
\label{sec:conclusion}

We have made a three-dimensional dust map of the giant molecular cloud Orion A by computing the best linear unbiased predictor (BLUP, Eq.~\ref{eq:blup}) of the extinction at every point on an arbitrarily dense lattice using extinctions for a set of observed stars in the region of this lattice.
The BLUP requires us to make very few assumptions about the statistical properties of the ISM.
It requires only that we know the covariance of the density at any two points, and that we have a model of the expected value of the density at any one point.
Beyond this knowledge of the first and second moments of the density, we need know nothing of its distribution.

In practice, however, we do not know the covariance of the extinctions at two points, and instead have a model of it, just as we have a model for the expected value.
Since we do not know the distribution of the dust density we are not able to use the method of maximum likelihood to optimize this model, but instead fit it to the data by choosing the parameter tuple that minimizes the LOOCV score.
It is also convenient to assume that the density field is stationary and isotropic, and that it is described by Kolmogorov's theory of turbulence within the inertial range set by the energy-injection and eneryg-dissipation scales.
Moreover, by assuming that the density is isotropic and homogeneous our model of the mean density has only one free parameter, $\mu_\density$, which we can recover in the course of making our predictions.

This agnosticism is both a strength and a weakness of our method, since it does not allow us to encode the fact that density is always non-negative.
Nonetheless, in experiments using synthetic data we find that predicted densities are, in practice, everywhere non-negative.
With observational data we find that predicted densities are, in practice, everywhere consistent with being non-negative.
Negative predicted densities are the result of spuriously high predicted extinctions regressing to the mean along the line of sight, meaning that, along a line of sight, the extinction field is not monotone.
Observational data also suffer from selection effects that cause the average observed extinction to decrease with distance \citep{sale_three-dimensional_2015}. The more heavily extinguished a star is the fainter it is and hence the less chance it has of appearing in any magnitude-limited catalogue.
These negative densities are therefore due to either a lack of data or to model mispecification.
In the latter case, either our assumptions about the statistics of the ISM fail to do it justice, or the data include sources with extinctions that are not due exclusively to the dust of the ISM (for example, the extinctions of young stellar objects is partially due to their dusty circumstellar discs and envelopes).
Such sources are likely to be present in any catalogue, especially in catalogues containing nebulae like those of Orion A, even after cleaning.

The choice of autocovariance function is crucial to the quality of any three-dimensional dust map.
We have used the physically motivated Kolmogorov-like function defined by \cite{sale_three-dimensional_2014}.
Other autocovariance functions may be used as substitutes so long as they have the appropriate properties, i.e. so long as they define a density field that is continuous and rough, in which case they seem to adequately describe the fractal behaviour of the ISM.
These requirements exclude the popular squared-exponential autocovariance function, which defines continuous but smooth random fields.
The use of a squared-exponential autocovariance function will result in smooth maps, the features of which will be washed out.
Nevertheless, the Kolmogorov-like function has a physically meaningful parameter tuple that library functions like the powered exponential and the Gneiting function do not.
However, by comparing the integral length scale of a substitute autocovariance function with that of the Kolmogorov-like autocovariance function we may find the physical length scale that corresponds to its characteristic length scale.
This physical length scale approximates the energy-injection scale, and hence gives physical meaning to the characteristic length scale of the substitute function.

The maps we have made using RJCE extinction data combined with line-of-sight distances inferred by \cite{bailer-jones_estimating_2021} broadly agree with those of \cite{rezaei_detailed_2020} and \cite{rezaei_three-dimensional_2022}.
In particular, they corroborate Rezaei Kh's claim that a foreground cloud exists at a distance of 350~kpc.
However, the maps we have made using StarHorse data \citep{anders_photo-astrometric_2019} fail even to identify the head of Orion A. Crucially, these maps fail validation, alerting us to the fact that they are untrustworthy.
The unfiltered StarHorse catalogues, though wonderful, must be used with care.
They contain many sources, such as young stellar objects, that violate the assumptions of its methodology, and which therefore have spurious extinction entries that are not flagged as such.

%% file: acknowledgements.tex
We gratefully acknowledge the support of the UKRI Science and Technology Facilities Council (grant ST/S000488/1).
We would like to thank Sara Rezaei Kh., for generously making the data of her 2022 paper available to us, and our anonymous reviewer, whose insightful comments significantly improved this publication.

This publication makes use of data products from the Two Micron All Sky Survey, which is a joint project of the University of Massachusetts and the Infrared Processing and Analysis Center/California Institute of Technology, funded by the National Aeronautics and Space Administration and the National Science Foundation.

This publication makes use of data products from the Wide-field Infrared Survey Explorer, which is a joint project of the University of California, Los Angeles, and the Jet Propulsion Laboratory/California Institute of Technology, and NEOWISE, which is a project of the Jet Propulsion Laboratory/California Institute of Technology. WISE and NEOWISE are funded by the National Aeronautics and Space Administration.

This publication makes use of data from the European Space Agency (ESA) mission {\it Gaia} (\url{https://www.cosmos.esa.int/gaia}), processed by the {\it Gaia} Data Processing and Analysis Consortium (DPAC, \url{https://www.cosmos.esa.int/web/gaia/dpac/consortium}). Funding for the DPAC has been provided by national institutions, in particular the institutions participating in the {\it Gaia} Multilateral Agreement.

%% file: data_availability.tex
The RJCE extinctions used in this paper were provided by Sara Rezaei Kh.
The StarHorse catalogue and the catalogue of Bailer-Jones are availabe at the \gaia Archive (\url{https://gea.esac.esa.int/archive/}) and the Gaia DR2 mirror archive (\url{gaia.aip.de}). Predictions of extinction and density were made using code written in Python. This code is available on request.

%% file: appendix_1.tex
\label{sec:appendix_1}

Our predictor for the density at a point, $\hat{\rho}(\bm{r})$ (Eq.~\ref{eq:predictor_density}), is the derivative of the BLUP of the extinction at that point, $\hat{A}(\bm{r})$.
To compute it, note that in Equation~\ref{eq:blup} only $\bm{\gamma}$ and $\bm{\sigma}$ are functions of $s$, so that
\begin{align}
  \hat{\rho}(\bm{r}) = \dfrac{\partial{}\bm{\gamma}^{\trans}}{\partial{}s}\bm{A} + \dfrac{\partial{}\bm{\sigma}^{\trans}}{\partial{}s}\bm{\Sigma}^{-1}(\bm{A} - \bm{\Gamma}^{\trans}\bm{A})
\end{align}
where
\begin{align}
  \dfrac{\partial{}\bm{\gamma}}{\partial{}s}
  = \bm{\Sigma}^{-1}{\bf{}\Phi}^{\trans}({\bf{}\Phi}\bm{\Sigma}^{-1}{\bf{}\Phi}^{\trans})^{-1}
\end{align}
and $[\partial{}\bm{\sigma}/\partial{}s]_{i} = \cov(\rho(\bm{r}), A(\bm{r}_{i}))$.
Alternatively, we may rewrite the BLUP of the extinction as $\hat{A}(\bm{r}) = \bm{\alpha}^{\trans}\bm{A}$ where 
\begin{align}
  \bm{\alpha} = (\bm{I} - \bm{\Gamma})\bm{\Sigma}^{-1}\bm{\sigma} + \bm{\gamma},
\end{align}
in which case our predictor of density may be rewritten as
\begin{align}
  \hat{\density}(\bm{r})
  &= \dfrac{\partial{}\bm{\alpha}^{\trans}}{\partial{}s}\bm{A}.
\end{align}
This involves the same partial derivative, $\partial\bm{\alpha}/\partial{}s$, that is involved in the expression for $\var(\density(\bm{r}) - \hat{\density}(\bm{r}))$ (Eq.~\ref{eq:prediction_interval_density}).
Note that
\begin{align}
  \dfrac{\partial{}\bm{\alpha}}{\partial{}s} = (\bm{I} - \bm{\Gamma})\bm{\Sigma}^{-1}\dfrac{\partial{}\bm{\sigma}}{\partial{}s} + \dfrac{\partial{}\bm{\gamma}}{\partial{}s},
\end{align}
which equips us with all the information we need to compute $\hat{\rho}(\bm{r})$ and its prediction interval (App.~\ref{sec:appendix_2}).

%% file: appendix_2.tex
\label{sec:appendix_2}

A prediction interval for $\hat{A}(\bm{r})$ is
\begin{align}
[\hat{A}(\bm{r}) - \lambda_{A}S_{A},
 \hat{A}(\bm{r}) + \lambda_{A}S_{A}]
\end{align}
where
\begin{align}
S_{A} := \sqrt{\var(A(\bm{r}) - \hat{A}(\bm{r}))}
\end{align}
and $\lambda_{A}$ is a critical value for the random variable $A(\bm{r}) - 
\hat{A}(\bm{r})$.
Because $\hat{A}(\bm{r})$ is unbiased the variance of $A(\bm{r}) - \hat{A}(\bm{r})$ is equal to the mean-square error of $\hat{A}(\bm{r})$ since
\begin{align}
  \mse(\hat{A}(\bm{r}))
  &= \var(A(\bm{r}) - \hat{A}(\bm{r})) + \bias^{2}(\hat{A}(\bm{r}))\\
  &= \var(A(\bm{r}) - \hat{A}(\bm{r})).
\end{align}
This mean-square error is the quantity minimized by the BLUP of the extinction, and is given by \cite{goldberger_best_1962} as
\begin{align}
  \label{eq:mse}
  \begin{split}
  \mse(\hat{A}(\bm{r}))
    &= \var(A(\bm{r})) - \bm{\sigma}^{\trans}\bm{\Sigma}^{-1}\bm{\sigma}\\
    &\qquad + (\bm{\phi} - {\bf{}\Phi}\bm{\Sigma}^{-1}\bm{\sigma})^{\trans}({\bf{}\Phi}\bm{\Sigma}^{-1}{\bf{}\Phi}^{\trans})^{-1}(\bm{\phi} - {\bf{}\Phi}\bm{\Sigma}^{-1}\bm{\sigma}).
  \end{split}
\end{align}
Similarly, a prediction interval for $\hat{\rho}(\bm{r})$ is
\begin{align}
[\hat{\rho}(\bm{r}) - \lambda_{\rho}S_{\rho},
 \hat{\rho}(\bm{r}) + \lambda_{\rho}S_{\rho}]
\end{align}
where
\begin{align}
S_{\rho} := \sqrt{\var(\rho(\bm{r}) - \hat{\rho}(\bm{r}))}
\end{align}
and $\lambda_{\rho}$ is a critical value for the random variable $\rho(\bm{r}) - 
\hat{\rho}(\bm{r})$.
Using Equations~\ref{eq:blup} and~\ref{eq:predictor_density} we find that this variance is
\begin{align}
  \var(\density(\bm{r}) - \hat{\density}(\bm{r}))
  &= \var(\density(\bm{r})) + \var(\hat{\density}(\bm{r})) - 2\cov(\density(\bm{r}), \hat{\density}(\bm{r}))\\
  \label{eq:prediction_interval_density}
  &= \sigma^{2}_{\densityfluctuations}
  + \dfrac{\partial{}\bm{\alpha}^{\trans}}{\partial{}s}\bm{\Sigma}\dfrac{\partial{}\bm{\alpha}}{\partial{}s} - 2\cov(\density(\bm{r}), \bm{A})\dfrac{\partial{}\bm{\alpha}}{\partial{}s}
\end{align}
where the derivative $\partial{}\bm{\alpha}/\partial{}s$ is computed in Appendix~\ref{sec:appendix_1}.
A confidence interval for $\mu_{\rho}$ is
\begin{align}
[\hat{\mu}_{\rho} - \lambda_{\mu_{\rho}}S_{\mu_{\rho}},
 \hat{\mu}_{\rho} + \lambda_{\mu_{\rho}}S_{\mu_{\rho}}]
\end{align}
where
\begin{align}
S_{\mu_{\rho}} := \sqrt{\var(\hat{\mu}_{\rho})}
\end{align}
and $\lambda_{\mu_{\rho}}$ is a critical value for $\hat{\mu_{\rho}}$.
This variance is
\begin{align}
\var(\hat{\mu}_{\rho}) = (\bm{\Phi}^{\trans}\bm{\Sigma}^{-1}\bm{\Phi})^{-1}.
\end{align}

To find the critical values $\lambda_{A}$, $\lambda_{\rho}$, and $\lambda_{\mu_{\density}}$ we must know the distributions of $A(\bm{r}) - \hat{A}(\bm{r})$, $\rho(\bm{r}) - \hat{\rho}(\bm{r})$, and $\hat{\mu}_{\rho}$ respectively.
But, as we have observed, the BLUP requires us to have knowledge only of the first and second moments of $(A(\bm{r}), \bm{A})$.
In fact, this is one of the principal benefits of the method.
If we do not know this distribution then we do not know the distributions of $A(\bm{r})- \hat{A}(\bm{r})$, $\rho(\bm{r}) - \hat{\rho}(\bm{r})$, and $\hat{\mu}_{\rho}$.
Instead, we can find limits for the prediction intervals using Chebyshev's inequality, which holds for all distributions.
In this situation, the $68$-$95$-$99.7$ rule for normal random variables does not hold.
Instead, the confidence levels $0.68, 0.95,$ and $0.997$ are associated with upper limits on the critical values of $1.77, 4.47,$ and $18.3$ rather than the usual values of 1, 2, and 3.

%% file: appendix_3.tex
\label{sec:appendix_3}

\begin{figure*}
  \centering
  \begin{minipage}[b]{.4\textwidth}
    \includegraphics[width=\columnwidth]{
      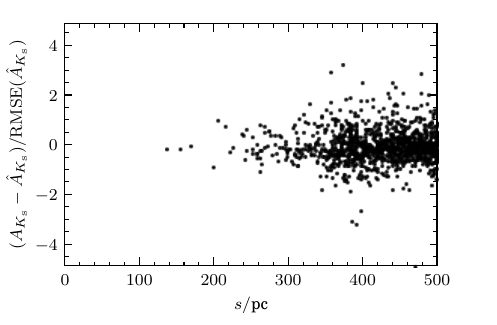
    }
    \includegraphics[width=\columnwidth]{
      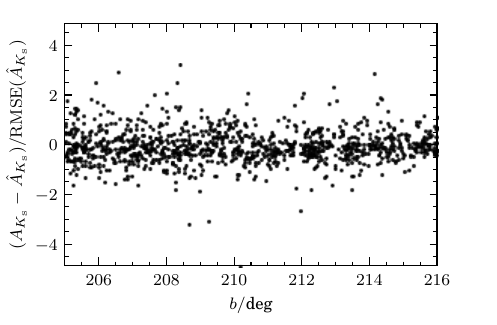
    }
    \includegraphics[width=\columnwidth]{
      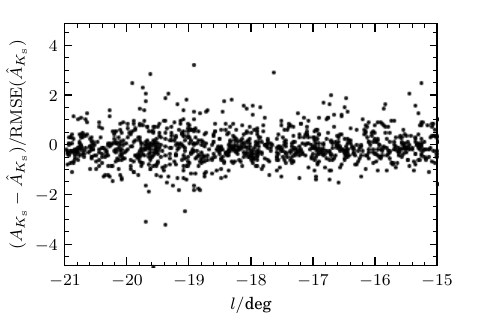
    }
  \caption{
  Colour excess and \gaia{} data:
  the standardized leave-one-out residuals of the predicted extinction.
  (For the sake of clarity these plots show a sample of size 1000, uniformly distributed in volume.)
  }
  \label{fig:rjce_data_loocv}
\end{minipage}
\qquad
\begin{minipage}[b]{.4\textwidth}
  \includegraphics[width=\columnwidth]{
    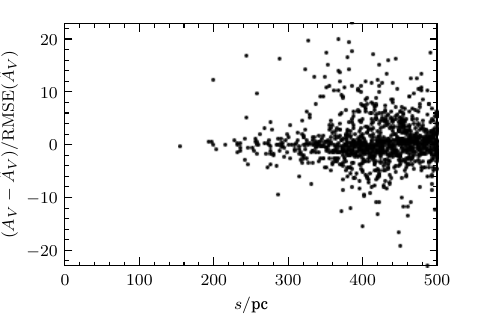
  }
  \includegraphics[width=\columnwidth]{
    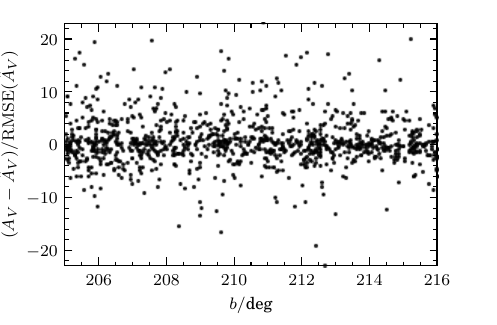
  }
  \includegraphics[width=\columnwidth]{
    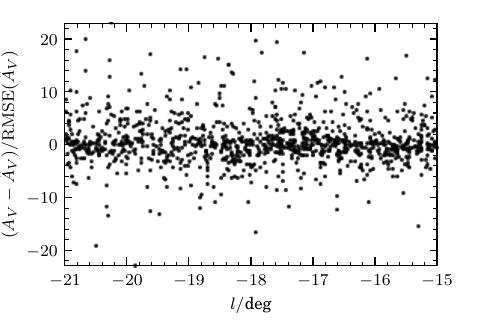
  }
  \caption{StarHorse data:
    the standardized leave-one-out residuals of the predicted extinction.
    (For the sake of clarity these plots show a sample of size 1000, uniformly distributed in volume.)
  }
\label{fig:starhorse_data_loocv}
\end{minipage}
\end{figure*}

Validation of the BLUP is crucial, and the process of validation helps weed out predictors that underperform.
The predictors of extinction that we construct using RJCE extinction data combined with line-of-sight distances inferred by \cite{bailer-jones_estimating_2021} pass validation.
But those that we construct using StarHorse data \citep{anders_photo-astrometric_2019} do not.
We plot the standardized leave-one-out residuals for these predictors in Figures~\ref{fig:rjce_data_loocv} and~\ref{fig:starhorse_data_loocv}.
We do not expect the residuals to have Gaussian distribution, but they must obey Chebyshev's inequality (see App.~\ref{sec:appendix_2}).

%% file: appendix_4.tex
\label{sec:appendix_4}

\begin{figure*}
  \centering
  \includegraphics{
    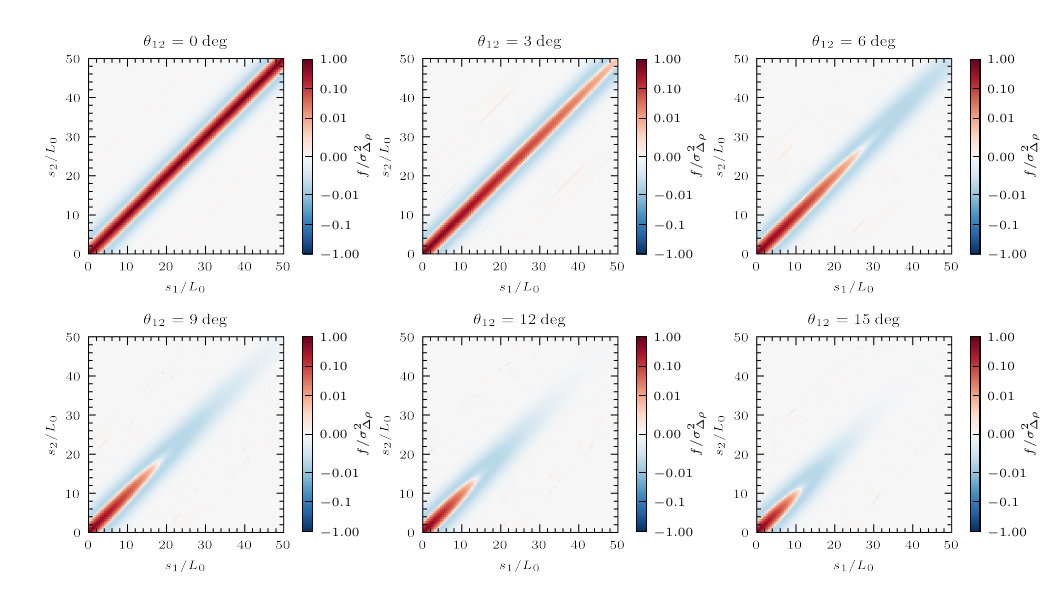
  }
  \caption{%
  Covariance of the density of the ISM at two points, $\bm{r}_{1}$ and $\bm{r}_{2}$, as a function of the distances, $s_{1}$ and $s_{2}$, and the angular separation, $\theta_{12}$ (the function $f$, eq.~\ref{eq:function_f}) for $\Omega = 1$ and $\gamma = 11/3$.
  In each panel the angular separation is fixed, and the distances allowed to vary.
  The covariance is shown on an arcsinh scale with linear width $0.001$.
  }
  \label{fig:function_f_a}
  \includegraphics{
    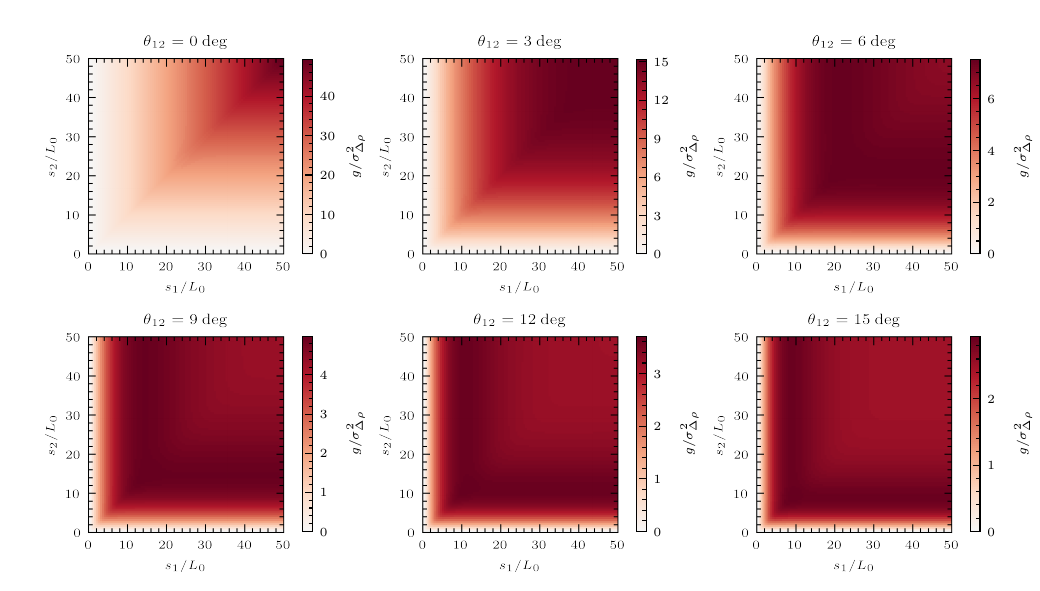
  }
  \caption{%
  Covariance of the extinction at two points, $\bm{r}_{1}$ and $\bm{r}_{2}$, as a function of the distances, $s_{1}$ and $s_{2}$, and the angular separation, $\theta_{12}$ (the function $g$, eq.~\ref{eq:function_g}) for $\Omega = 1$ and $\gamma = 11/3$.
  In each panel the angular separation is fixed, and the distances allowed to vary.
  }
  \label{fig:function_g_a}
\end{figure*}

We have introduced the functions $f$ and $g$ (\ref{eq:function_f} and \ref{eq:function_g}), which give the covariance of two points drawn from the density and extinction fields in terms of their line-of-sight distances and angular separation.
We plot these functions in Figures~\ref{fig:function_f_a}, \ref{fig:function_g_a}, and \ref{fig:functions_fg} for $\Omega = 1$ and $\gamma = 11/3$ under the assumption of unit variance, $\sigma_{\densityfluctuations}^{2}$, and unit length scale, $L_{0}$.
For a given angular separation, $\theta_{12}$, the functions $f$ and $g$ are both symmetric in $s_{1}$ and $s_{2}$.
In the case that $\theta = 0~\mathrm{deg}$, the diagonals give the variance at a line-of-sight distance $s_{1} = s_{2}$, while in the case of $\theta_{12} > 0~\mathrm{deg}$ the origins give the variance, and the diagonals give the covariance at a common distance.

\begin{figure*}
  \centering
  \includegraphics{
    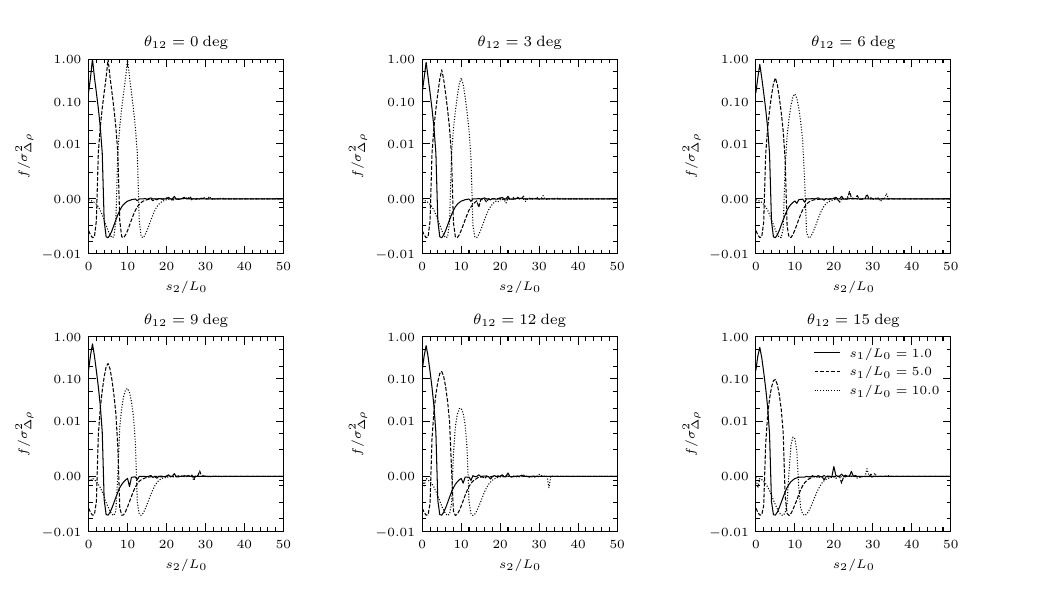
  }
  \caption{%
  Covariance of the density of the ISM at two points, $\bm{r}_{1}$ and $\bm{r}_{2}$, as a function of the distances, $s_{1}$ and $s_{2}$, and the angular separation, $\theta_{12}$ (the function $f$, eq.~\ref{eq:function_f}) for $\Omega = 1$ and $\gamma = 11/3$.
  In each panel the distance $s_{1}$ and the angular separation are fixed, and the distance $s_{2}$ allowed to vary.
  The covariance is shown on an arcsinh scale with linear width $0.001$.
  }
  \label{fig:function_f_b}
  \includegraphics{
    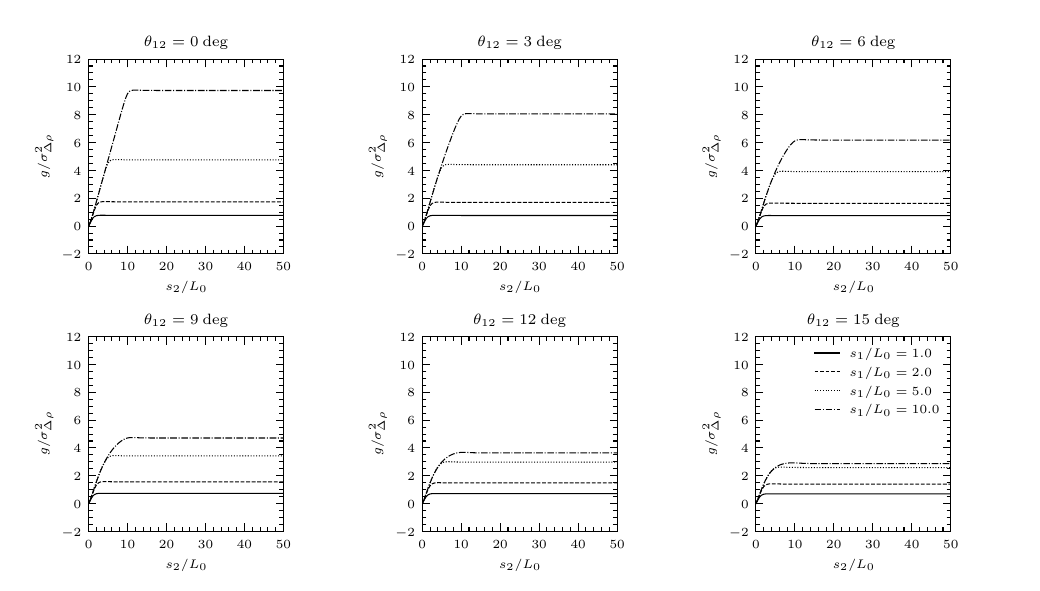
  }
  \caption{%
  Covariance of the extinction at two points, $\bm{r}_{1}$ and $\bm{r}_{2}$, as a function of the distances, $s_{1}$ and $s_{2}$, and the angular separation, $\theta_{12}$ (the function $g$, eq.~\ref{eq:function_g}) for $\Omega = 1$ and $\gamma = 11/3$.
  In each panel the distance $s_{1}$ and the angular separation are fixed, and the distance $s_{2}$ allowed to vary.
  }
  \label{fig:function_g_b}
\end{figure*}

\begin{figure}
  \centering
  \includegraphics{
    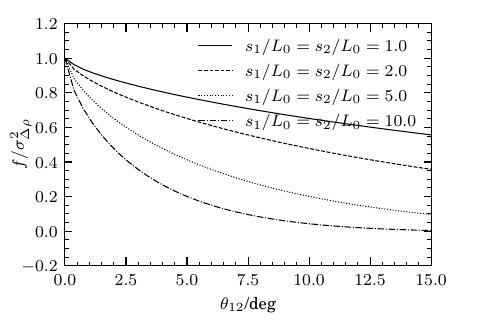
  }
  \includegraphics{
    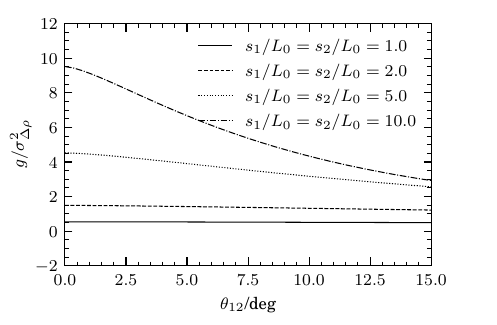
  }
  \caption{%
  Covariance of the density of the ISM (top, the function $f$, eq.~\ref{eq:function_f}) and of the extinction (bottom, the function $g$, eq.~\ref{eq:function_g}) for $\Omega = 1$, $\gamma = 11/3$, and unit length scale, $L_{0}$.
  See also Figures~\ref{fig:function_f_a}, \ref{fig:function_f_b}, \ref{fig:function_g_a}, \ref{fig:function_g_b} and their captions.
  In each panel the distances $s_{1}$ and $s_{2}$ are equal and the angular separation, $\theta_{12}$, allowed to vary.
  }
  \label{fig:functions_fg}
\end{figure}

For all values of $\theta$, the function $f$ is peaked on the diagonal, and approximately zero far from it.
For $\theta = 0~\mathrm{deg}$ it is constant along the diagonal,
and for $\theta > 0~\mathrm{deg}$ it decreases along the diagonal from a maximum at the origin.
The greater the angular separation, the faster it decreases.
This behaviour is seen more clearly seen in Figure~\ref{fig:function_f_b}, which shows $f$ for fixed $s_{1}$ and $\theta_{12}$ and variable $s_{2}$.

For $\theta = 0~\mathrm{deg}$, the function $g$ is linear in $s_{1}$ and constant in $s_{2}$ above the diagonal.
Below the diagonal it is constant in $s_{1}$ and linear in $s_{2}$.
This is to say that, along a line of sight, the extinction at a point is strongly correlated with the extinction at a second point more distant since that extinction must be at least as great, and decreasingly correlated with the extinction at a point less distant, since that extinction must be smaller.
There is a smooth transition across the diagonal, more clearly seen in Figure~\ref{fig:function_g_b}, the width of which is approximately $L_{0}$.
This smooth transition is due to the fact that densities, and hence extinctions, are strongly correlated at distances of order $L_{0}$ or less, whether the second point is more or less distant than the first.
For $\theta > 0~\mathrm{deg}$, the function $g$ above the diagonal is linear in $s_{1}$ up to some critical value and constant in $s_{1}$ beyond that critical value, whilst being constant in $s_{2}$.
Below the diagonal the function $g$ is constant in $s_{1}$, whilst being linear in $s_{2}$ up to the critical value and constant in $s_{2}$ beyond the critical value.
The greater the angular separation, the smaller the linear regime.
Along distinct lines of sight, the extinction at a point is less strongly correlated with the extinction at a second point more distant since that extinction need not be as great and decreasingly correlated with the extinction at a point less distant since that extinction must again be zero at the origin.
Again, there is a smooth transition across the diagonal, more clearly in Figure~\ref{fig:function_g_b}, due to the fact that extinctions are strongly correlated at distances of order $L_{0}$ or less.

\subsection{Alternative covariance functions}

\begin{figure}
  \centering
  \includegraphics{
    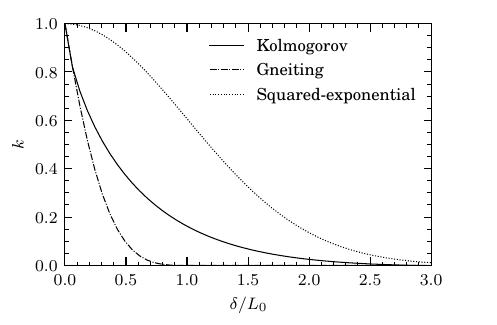
  }
  \caption{%
    The squared-exponential autocovariance function, $k_{\text{PE}}$ (eq.~\ref{eq:powered_exponential_function}, $\alpha = 2$), Gneiting autocovariance function, $k_{\text{G}}$ (eq.~\ref{eq:gneiting_function}), and Kolmogorov-like autocovariance function, $k_{\densityfluctuations}$ (eqs~\ref{eq:autocorrelation_model}, \ref{eq:psd_model}), for the case of unit variance and unit characteristic length scale.
  The Kolmogorov-like autocovariance function has power-law indices 
  $\Omega = 1$ and $\gamma = 11/3$.
  }
  \label{fig:autocov_functions}
\end{figure}

We have also discussed the use of autocovariance functions that have no direct physical motivation, and in particular the use of the squared-exponential and Gneiting functions (Sec.~\ref{sec:discussion}).
In Figure~\ref{fig:autocov_functions} we show the graphs of these, alongside the Kolmogorov-like autocovariance function, all for the case of unit variance and unit characteristic length scale.

%% file: main.bbl
\begin{thebibliography}{}
\makeatletter
\relax
\def\mn@urlcharsother{\let\do\@makeother \do\$\do\&\do\#\do\^\do\_\do\%\do\~}
\def\mn@doi{\begingroup\mn@urlcharsother \@ifnextchar [ {\mn@doi@}
  {\mn@doi@[]}}
\def\mn@doi@[#1]#2{\def\@tempa{#1}\ifx\@tempa\@empty \href
  {http://dx.doi.org/#2} {doi:#2}\else \href {http://dx.doi.org/#2} {#1}\fi
  \endgroup}
\def\mn@eprint#1#2{\mn@eprint@#1:#2::\@nil}
\def\mn@eprint@arXiv#1{\href {http://arxiv.org/abs/#1} {{\tt arXiv:#1}}}
\def\mn@eprint@dblp#1{\href {http://dblp.uni-trier.de/rec/bibtex/#1.xml}
  {dblp:#1}}
\def\mn@eprint@#1:#2:#3:#4\@nil{\def\@tempa {#1}\def\@tempb {#2}\def\@tempc
  {#3}\ifx \@tempc \@empty \let \@tempc \@tempb \let \@tempb \@tempa \fi \ifx
  \@tempb \@empty \def\@tempb {arXiv}\fi \@ifundefined
  {mn@eprint@\@tempb}{\@tempb:\@tempc}{\expandafter \expandafter \csname
  mn@eprint@\@tempb\endcsname \expandafter{\@tempc}}}

\bibitem[\protect\citeauthoryear{Adler}{Adler}{1981}]{adler_geometry_1981}
Adler R.~J.,  1981, The geometry of random fields.
J. Wiley, Chichester

\bibitem[\protect\citeauthoryear{Anders et~al.,}{Anders
  et~al.}{2019}]{anders_photo-astrometric_2019}
Anders F.,  et~al., 2019, Astronomy \& Astrophysics, 628, A94

\bibitem[\protect\citeauthoryear{Anders et~al.,}{Anders
  et~al.}{2021}]{anders_photo-astrometric_2021}
Anders F.,  et~al., 2021, arXiv:2111.01860

\bibitem[\protect\citeauthoryear{Arenou, Grenon  \& Gomez}{Arenou
  et~al.}{1992}]{arenou_tridimensional_1992}
Arenou F.,  Grenon M.,   Gomez A.,  1992, Astronomy and Astrophysics, 258, 104

\bibitem[\protect\citeauthoryear{Astraatmadja \& Bailer-Jones}{Astraatmadja \&
  Bailer-Jones}{2016a}]{astraatmadja_estimating_2016a}
Astraatmadja T.~L.,  Bailer-Jones C. A.~L.,  2016a, \mn@doi [The Astrophysical
  Journal] {10.3847/0004-637X/832/2/137}, 832, 137

\bibitem[\protect\citeauthoryear{Astraatmadja \& Bailer-Jones}{Astraatmadja \&
  Bailer-Jones}{2016b}]{astraatmadja_estimating_2016b}
Astraatmadja T.~L.,  Bailer-Jones C. A.~L.,  2016b, \mn@doi [The Astrophysical
  Journal] {10.3847/1538-4357/833/1/119}, 833, 119

\bibitem[\protect\citeauthoryear{Bailer-Jones}{Bailer-Jones}{2011}]{bailer-jones_bayesian_2011}
Bailer-Jones C. A.~L.,  2011, Monthly Notices of the Royal Astronomical
  Society, 411, 435

\bibitem[\protect\citeauthoryear{Bailer-Jones}{Bailer-Jones}{2015}]{bailer-jones_estimating_2015}
Bailer-Jones C. A.~L.,  2015, \mn@doi [Publications of the Astronomical Society
  of the Pacific] {10.1086/683116}, 127, 994

\bibitem[\protect\citeauthoryear{Bailer-Jones, Rybizki, Fouesneau, Mantelet  \&
  Andrae}{Bailer-Jones et~al.}{2018}]{bailer-jones_estimating_2018}
Bailer-Jones C. A.~L.,  Rybizki J.,  Fouesneau M.,  Mantelet G.,   Andrae R.,
  2018, \mn@doi [The Astronomical Journal] {10.3847/1538-3881/aacb21}, 156, 58

\bibitem[\protect\citeauthoryear{Bailer-Jones, Rybizki, Fouesneau, Demleitner
  \& Andrae}{Bailer-Jones et~al.}{2021}]{bailer-jones_estimating_2021}
Bailer-Jones C. A.~L.,  Rybizki J.,  Fouesneau M.,  Demleitner M.,   Andrae R.,
   2021, \mn@doi [The Astronomical Journal] {10.3847/1538-3881/abd806}, 161,
  147

\bibitem[\protect\citeauthoryear{Berry et~al.,}{Berry
  et~al.}{2012}]{berry_milky_2012}
Berry M.,  et~al., 2012, The Astrophysical Journal, 757, 166

\bibitem[\protect\citeauthoryear{Binney et~al.,}{Binney
  et~al.}{2014}]{binney_galactic_2014}
Binney J.,  et~al., 2014, Monthly Notices of the Royal Astronomical Society,
  439, 1231

\bibitem[\protect\citeauthoryear{Bouy, Alves, Bertin, Sarro  \& Barrado}{Bouy
  et~al.}{2014}]{bouy_orion_2014}
Bouy H.,  Alves J.,  Bertin E.,  Sarro L.~M.,   Barrado D.,  2014, Astronomy \&
  Astrophysics, 564, A29

\bibitem[\protect\citeauthoryear{Burnett \& Binney}{Burnett \&
  Binney}{2010}]{burnett_stellar_2010}
Burnett B.,  Binney J.,  2010, Monthly Notices of the Royal Astronomical
  Society, 407, 339

\bibitem[\protect\citeauthoryear{Cardelli, Clayton  \& Mathis}{Cardelli
  et~al.}{1989}]{cardelli_relationship_1989}
Cardelli J.~A.,  Clayton G.~C.,   Mathis J.~S.,  1989, The Astrophysical
  Journal, 345, 245

\bibitem[\protect\citeauthoryear{Carpenter}{Carpenter}{2000}]{carpenter_2mass_2000}
Carpenter J.~M.,  2000, The Astronomical Journal, 120, 3139

\bibitem[\protect\citeauthoryear{Chen et~al.,}{Chen
  et~al.}{2014}]{chen_three-dimensional_2014}
Chen B.~Q.,  et~al., 2014, Monthly Notices of the Royal Astronomical Society,
  443, 1192

\bibitem[\protect\citeauthoryear{Coles \& Barrow}{Coles \&
  Barrow}{1987}]{coles_nongaussian_1987}
Coles P.,  Barrow J.~D.,  1987, Monthly Notices of the Royal Astronomical
  Society, 228, 407

\bibitem[\protect\citeauthoryear{Cressie}{Cressie}{1993}]{cressie_statistics_1993}
Cressie N.,  1993, Statistics for spatial data.
Wiley series in probability and statistics, J. Wiley, New York

\bibitem[\protect\citeauthoryear{Dharmawardena, Bailer-Jones, Fouesneau  \&
  Foreman-Mackey}{Dharmawardena
  et~al.}{2022}]{dharmawardena_three-dimensional_2022}
Dharmawardena T.~E.,  Bailer-Jones C. A.~L.,  Fouesneau M.,   Foreman-Mackey
  D.,  2022, Astronomy \& Astrophysics, 658

\bibitem[\protect\citeauthoryear{Draine}{Draine}{2011}]{draine_physics_2011}
Draine B.~T.,  2011, Physics of the interstellar and intergalactic medium.
Princeton University Press, Princeton, N.J

\bibitem[\protect\citeauthoryear{Drimmel, Cabrera-Lavers  \&
  López-Corredoira}{Drimmel et~al.}{2003}]{drimmel_three-dimensional_2003}
Drimmel R.,  Cabrera-Lavers A.,   López-Corredoira M.,  2003, Astronomy \&
  Astrophysics, 409, 205

\bibitem[\protect\citeauthoryear{En{\ss}lin \& Frommert}{En{\ss}lin \&
  Frommert}{2011}]{enslin_reconstruction_2011}
En{\ss}lin T.~A.,  Frommert M.,  2011, Physical Review D, 83, 105014

\bibitem[\protect\citeauthoryear{En{\ss}lin \& Weig}{En{\ss}lin \&
  Weig}{2010}]{enslin_inference_2010}
En{\ss}lin T.~A.,  Weig C.,  2010, Physical Review E, 82, 051112

\bibitem[\protect\citeauthoryear{{Gaia Collaboration}}{{Gaia
  Collaboration}}{2018}]{gaia_collaboration_gaia_2018}
{Gaia Collaboration} 2018, Astronomy \& Astrophysics, 616, A1

\bibitem[\protect\citeauthoryear{Gneiting}{Gneiting}{2002}]{gneiting_compactly_2002}
Gneiting T.,  2002, Journal of Multivariate Analysis, 83, 493

\bibitem[\protect\citeauthoryear{Goldberger}{Goldberger}{1962}]{goldberger_best_1962}
Goldberger A.~S.,  1962, Journal of the American Statistical Association, 57

\bibitem[\protect\citeauthoryear{Green et~al.,}{Green
  et~al.}{2014}]{green_measuring_2014}
Green G.~M.,  et~al., 2014, The Astrophysical Journal, 783, 114

\bibitem[\protect\citeauthoryear{Green et~al.,}{Green
  et~al.}{2015}]{green_three-dimensional_2015}
Green G.~M.,  et~al., 2015, The Astrophysical Journal, 810, 25

\bibitem[\protect\citeauthoryear{Green et~al.,}{Green
  et~al.}{2018}]{green_galactic_2018}
Green G.~M.,  et~al., 2018, Monthly Notices of the Royal Astronomical Society,
  478, 651

\bibitem[\protect\citeauthoryear{Green, Schlafly, Zucker, Speagle  \&
  Finkbeiner}{Green et~al.}{2019}]{green_3d_2019}
Green G.~M.,  Schlafly E.,  Zucker C.,  Speagle J.~S.,   Finkbeiner D.,  2019,
  The Astrophysical Journal, 887, 93

\bibitem[\protect\citeauthoryear{Gro{\ss}schedl et~al.,}{Gro{\ss}schedl
  et~al.}{2018}]{grosschedl_3d_2018}
Gro{\ss}schedl J.~E.,  et~al., 2018, 619

\bibitem[\protect\citeauthoryear{Gro{\ss}schedl et~al.,}{Gro{\ss}schedl
  et~al.}{2019}]{grosschedl_vision_2019}
Gro{\ss}schedl J.~E.,  et~al., 2019, Astronomy \& Astrophysics, 622, A149

\bibitem[\protect\citeauthoryear{Kounkel et~al.,}{Kounkel
  et~al.}{2018}]{kounkel_apogee-2_2018}
Kounkel M.,  et~al., 2018, The Astronomical Journal, 156, 84

\bibitem[\protect\citeauthoryear{Lada, Lada, Clemens  \& Bally}{Lada
  et~al.}{1994}]{lada_dust_1994}
Lada C.~J.,  Lada E.~A.,  Clemens D.~P.,   Bally J.,  1994, The Astrophysical
  Journal, 429, 694

\bibitem[\protect\citeauthoryear{Lallement}{Lallement}{2015}]{lallement_3d_2015}
Lallement R.,  2015, Journal of Physics: Conference Series, 577, 012016

\bibitem[\protect\citeauthoryear{Lallement, Vergely, Valette, Puspitarini, Eyer
   \& Casagrande}{Lallement et~al.}{2014}]{lallement_3d_2014}
Lallement R.,  Vergely J.-L.,  Valette B.,  Puspitarini L.,  Eyer L.,
  Casagrande L.,  2014, Astronomy \& Astrophysics, 561, A91

\bibitem[\protect\citeauthoryear{Lallement et~al.,}{Lallement
  et~al.}{2018}]{lallement_three-dimensional_2018}
Lallement R.,  et~al., 2018, Astronomy \& Astrophysics, 616, A132

\bibitem[\protect\citeauthoryear{Lallement, Babusiaux, Vergely, Katz, Arenou,
  Valette, Hottier  \& Capitanio}{Lallement
  et~al.}{2019}]{lallement_gaia-2mass_2019}
Lallement R.,  Babusiaux C.,  Vergely J.~L.,  Katz D.,  Arenou F.,  Valette B.,
   Hottier C.,   Capitanio L.,  2019, Astronomy \& Astrophysics, 625, A135

\bibitem[\protect\citeauthoryear{Leike \& En{\ss}lin}{Leike \&
  En{\ss}lin}{2019}]{leike_charting_2019}
Leike R.~H.,  En{\ss}lin T.~A.,  2019, Astronomy and Astrophysics, 631, A32

\bibitem[\protect\citeauthoryear{Leike, Glatzle  \& En{\ss}lin}{Leike
  et~al.}{2020}]{leike_resolving_2020}
Leike R.~H.,  Glatzle M.,   En{\ss}lin T.~A.,  2020, Astronomy and
  Astrophysics, 639, A138

\bibitem[\protect\citeauthoryear{Lombardi}{Lombardi}{2009}]{lombardi_nicest_2009}
Lombardi M.,  2009, Astronomy \& Astrophysics, 493, 735

\bibitem[\protect\citeauthoryear{Lombardi \& Alves}{Lombardi \&
  Alves}{2001}]{lombardi_mapping_2001}
Lombardi M.,  Alves J.,  2001, Astronomy and Astrophysics, 377, 1023

\bibitem[\protect\citeauthoryear{Majewski, Zasowski  \& Nidever}{Majewski
  et~al.}{2011}]{majewski_lifting_2011}
Majewski S.~R.,  Zasowski G.,   Nidever D.~L.,  2011, The Astrophysical
  Journal, 739

\bibitem[\protect\citeauthoryear{Marshall, Robin, Reylé, Schultheis  \&
  Picaud}{Marshall et~al.}{2006}]{marshall_modelling_2006}
Marshall D.~J.,  Robin A.~C.,  Reylé C.,  Schultheis M.,   Picaud S.,  2006,
  Astronomy and Astrophysics, 453, 635

\bibitem[\protect\citeauthoryear{Megeath et~al.,}{Megeath
  et~al.}{2012}]{megeath_spitzer_2012}
Megeath S.~T.,  et~al., 2012, The Astronomical Journal, 144, 192

\bibitem[\protect\citeauthoryear{Megeath et~al.,}{Megeath
  et~al.}{2015}]{megeath_spitzer_2015}
Megeath S.~T.,  et~al., 2015, The Astronomical Journal, 151, 5

\bibitem[\protect\citeauthoryear{Meingast et~al.,}{Meingast
  et~al.}{2016}]{meingast_vision_2016}
Meingast S.,  et~al., 2016, Astronomy \& Astrophysics, 587, A153

\bibitem[\protect\citeauthoryear{Meingast, Alves  \& Lombardi}{Meingast
  et~al.}{2018}]{meingast_vision_2018}
Meingast S.,  Alves J.,   Lombardi M.,  2018, Astronomy \& Astrophysics, 614,
  A65

\bibitem[\protect\citeauthoryear{Ostriker, Stone  \& Gammie}{Ostriker
  et~al.}{2001}]{ostriker_density_2001}
Ostriker E.~C.,  Stone J.~M.,   Gammie C.~F.,  2001, The Astrophysical Journal,
  546, 980

\bibitem[\protect\citeauthoryear{Queiroz et~al.,}{Queiroz
  et~al.}{2018}]{queiroz_starhorse_2018}
Queiroz A. B.~A.,  et~al., 2018, Monthly Notices of the Royal Astronomical
  Society, 476

\bibitem[\protect\citeauthoryear{Rasmussen}{Rasmussen}{2006}]{rasmussen_gaussian_2006}
Rasmussen C.~E.,  2006, Gaussian processes for machine learning.
MIT, Cambridge, Mass.

\bibitem[\protect\citeauthoryear{Rezaei~Kh. \& Kainulainen}{Rezaei~Kh. \&
  Kainulainen}{2022}]{rezaei_three-dimensional_2022}
Rezaei~Kh. S.,  Kainulainen J.,  2022, The Astrophysical Journal Letters, 930,
  L22

\bibitem[\protect\citeauthoryear{Rezaei~Kh, Bailer-Jones, Hanson  \&
  Fouesneau}{Rezaei~Kh et~al.}{2017}]{rezaei_inferring_2017}
Rezaei~Kh S.,  Bailer-Jones C. a.~L.,  Hanson R.~J.,   Fouesneau M.,  2017,
  Astronomy \& Astrophysics, 598, A125

\bibitem[\protect\citeauthoryear{Rezaei~Kh, Bailer-Jones, Schlafly  \&
  Fouesneau}{Rezaei~Kh et~al.}{2018}]{rezaei_three-dimensional_2018}
Rezaei~Kh S.,  Bailer-Jones C. a.~L.,  Schlafly E.~F.,   Fouesneau M.,  2018,
  Astronomy \& Astrophysics, 616, A44

\bibitem[\protect\citeauthoryear{Rezaei~Kh, Bailer-Jones, Soler  \&
  Zari}{Rezaei~Kh et~al.}{2020}]{rezaei_detailed_2020}
Rezaei~Kh S.,  Bailer-Jones C. A.~L.,  Soler J.~D.,   Zari E.,  2020, Astronomy
  \& Astrophysics, 643, A151

\bibitem[\protect\citeauthoryear{Sale}{Sale}{2012}]{sale_3d_2012}
Sale S.~E.,  2012, Monthly Notices of the Royal Astronomical Society, 427, 2119

\bibitem[\protect\citeauthoryear{Sale}{Sale}{2015}]{sale_three-dimensional_2015}
Sale S.~E.,  2015, 452, 2960

\bibitem[\protect\citeauthoryear{Sale \& Magorrian}{Sale \&
  Magorrian}{2014}]{sale_three-dimensional_2014}
Sale S.~E.,  Magorrian J.,  2014, Monthly notices of the Royal Astronomical
  Society, 445

\bibitem[\protect\citeauthoryear{Sale et~al.,}{Sale
  et~al.}{2014}]{sale_3d_2014}
Sale S.~E.,  et~al., 2014, Monthly Notices of the Royal Astronomical Society,
  443, 2907

\bibitem[\protect\citeauthoryear{Santiago et~al.,}{Santiago
  et~al.}{2016}]{santiago_spectro-photometric_2016}
Santiago B.~X.,  et~al., 2016, Astronomy \& Astrophysics, 585, A42

\bibitem[\protect\citeauthoryear{Schlafly et~al.,}{Schlafly
  et~al.}{2015}]{schlafly_three-dimensional_2015}
Schlafly E.~F.,  et~al., 2015, 799, 116

\bibitem[\protect\citeauthoryear{Skrutskie et~al.,}{Skrutskie
  et~al.}{2006}]{skrutskie_two_2006}
Skrutskie M.~F.,  et~al., 2006, The Astronomical Journal, 131, 1163

\bibitem[\protect\citeauthoryear{Tennekes}{Tennekes}{1972}]{tennekes_first_1972}
Tennekes H.,  1972, A first course in turbulence.
{MIT Press}

\bibitem[\protect\citeauthoryear{Vergely, Freire~Ferrero, Siebert  \&
  Valette}{Vergely et~al.}{2001}]{vergely_nai_2001}
Vergely J.-L.,  Freire~Ferrero R.,  Siebert A.,   Valette B.,  2001, Astronomy
  and Astrophysics, v.366, p.1016-1034 (2001), 366, 1016

\bibitem[\protect\citeauthoryear{Vergely, Valette, Lallement  \&
  Raimond}{Vergely et~al.}{2010}]{vergely_spatial_2010}
Vergely J.-L.,  Valette B.,  Lallement R.,   Raimond S.,  2010, Astronomy and
  Astrophysics, 518, A31

\bibitem[\protect\citeauthoryear{Wang, Pleiss, Gardner, Tyree, Weinberger  \&
  Wilson}{Wang et~al.}{2019}]{wang2019}
Wang K.~A.,  Pleiss G.,  Gardner J.~R.,  Tyree S.,  Weinberger K.~Q.,   Wilson
  A.~G.,  2019, Advances in Neural Information Processing Systems, 32

\bibitem[\protect\citeauthoryear{Wright et~al.,}{Wright
  et~al.}{2010}]{wright_wide-field_2010}
Wright E.~L.,  et~al., 2010, The Astronomical Journal, 140, 1868

\makeatother
\end{thebibliography}
